\documentclass[10pt,letterpaper]{article}
\usepackage[T1]{fontenc}
\usepackage[utf8]{inputenc}
\usepackage{mathptmx}
\usepackage{amsmath,amssymb,amsthm,bm}
\usepackage[letterpaper,textwidth=5.5in,textheight=9in,top=0.9in]{geometry}
\usepackage{graphicx,booktabs,array,multirow}
\usepackage{float}
\floatstyle{ruled}
\newfloat{algorithm}{tbp}{loa}
\floatname{algorithm}{Algorithm}
\newcounter{algorithmline}
\newcounter{algindent}
\newenvironment{algorithmic}[1][]{\setcounter{algindent}{0}\begin{list}{\arabic{algorithmline}:}{\usecounter{algorithmline}\setlength{\leftmargin}{1.8em}\setlength{\labelwidth}{1.4em}\setlength{\labelsep}{0.4em}\setlength{\itemsep}{0pt}\setlength{\parsep}{0pt}\setlength{\topsep}{3pt}}}{\end{list}}
\newcommand{\State}{\item\hspace*{\value{algindent}em}}
\newcommand{\Require}{\item[]\hspace*{-1.8em}\textbf{Require:} }
\newcommand{\For}[1]{\State\textbf{for} #1 \textbf{do}\stepcounter{algindent}}
\newcommand{\If}[1]{\State\textbf{if} #1 \textbf{then}\stepcounter{algindent}}
\newcommand{\Else}{\addtocounter{algindent}{-1}\State\textbf{else}\stepcounter{algindent}}
\newcommand{\ElsIf}[1]{\addtocounter{algindent}{-1}\State\textbf{else if} #1 \textbf{then}\stepcounter{algindent}}
\newcommand{\EndIf}{\addtocounter{algindent}{-1}\State\textbf{end if}}
\newcommand{\EndFor}{\addtocounter{algindent}{-1}\State\textbf{end for}}
\usepackage{enumitem,titlesec,microtype}
\usepackage[section]{placeins}
\usepackage[hidelinks]{hyperref}
\hypersetup{pdftitle={Not Every Token Needs Scoring: Query-Efficient LLM Watermark Localization},pdfauthor={Liyuan Wang, Xinsong Ma, Ziye Ma}}
\titleformat{\section}{\large\scshape}{\thesection}{0.8em}{}
\titleformat{\subsection}{\normalsize\scshape}{\thesubsection}{0.8em}{}
\titleformat{\subsubsection}{\normalsize\bfseries}{\thesubsubsection}{0.8em}{}
\titlespacing*{\section}{0pt}{13pt}{6pt}
\titlespacing*{\subsection}{0pt}{10pt}{4pt}
\setlist[itemize]{leftmargin=1.8em,itemsep=3pt,topsep=3pt}
\setlist[enumerate]{leftmargin=1.8em,itemsep=3pt,topsep=3pt}
\newtheorem{definition}{Definition}[section]
\newtheorem{theorem}[definition]{Theorem}
\newtheorem{proposition}{Proposition}

\newtheorem{lemma}{Lemma}
\newtheorem{appendixtheorem}{Theorem}
\newcommand{\E}{\mathbb{E}}
\newcommand{\1}{\mathbf{1}}
\newcommand{\Hnull}{\mathcal{H}_0}
\newcommand{\Halt}{\mathcal{H}_1}
\newcommand{\Shat}{\widehat{S}}
\newcommand{\Aw}{\mathcal{A}_{W}}
\newcommand{\para}[1]{\textbf{#1}\enspace}
\newcommand{\figpaper}[4]{\begin{figure}[htbp]\centering\includegraphics[width=#2\linewidth]{figures/#1.pdf}\caption{#3}\label{#4}\end{figure}}
\newcommand{\figpaperhere}[4]{\begin{figure}[H]\centering\includegraphics[width=#2\linewidth]{figures/#1.pdf}\caption{#3}\label{#4}\end{figure}}
\DeclareMathOperator*{\argmax}{arg\,max}
\DeclareMathOperator*{\argmin}{arg\,min}
\DeclareMathOperator{\clip}{clip}
\DeclareMathOperator{\FDR}{FDR}
\DeclareMathOperator{\FDP}{FDP}
\DeclareMathOperator{\Recall}{Recall}
\DeclareMathOperator{\TQR}{TQR}
\DeclareMathOperator{\Gain}{Gain}
\title{\vspace{-1.5em}\Large\textsc{Query-efficient scan-and-search watermark localization}}
\author{
\begin{tabular}{@{}c@{\quad}c@{\quad}c@{}}
Liyuan Wang &
Xinsong Ma\thanks{Corresponding author.} &
Ziye Ma \\[4pt]
\scriptsize\texttt{liyuawang9-c@my.cityu.edu.hk} &
\scriptsize\texttt{maxinsong1018@gmail.com} &
\scriptsize\texttt{ziyema@cityu.edu.hk}
\end{tabular}\\[8pt]
\normalsize Department of Computer Science\\
\normalsize City University of Hong Kong
}
\date{}
\begin{document}
\maketitle
\vspace{-1.6em}
\begin{abstract}
Recently, the detection of AI-generated content has been of particular importance in various fields of our society. For accountability purposes, many large language models (LLMs) have implicitly watermarked their outputs. Localizing such watermarks requires a piece of text to undergo an additional verification process. Existing localization methods typically rely on dense token-level scoring, making verification costly even when an LLM-generated watermark only sparsely appears. To tackle this problem, we design a new adaptive allocation strategy that first maintains broad document coverage and then progressively concentrates the detection budget on regions that exhibit stronger watermark evidence or are more likely to contain watermarked content. We then transform the statistical signals from the selected regions into e-values and employ the proposed token-weighted e-BH procedure to determine the localized watermark regions. The resulting framework is called BELL (Budgeted E-evidence Localization for LLM Watermarks). Theoretically, BELL can control the token-weighted false discovery rate (FDR) without independence assumptions across regions. Finally, extensive experimental results demonstrate that BELL consistently outperforms competing methods across multiple LLMs, datasets, and watermarking schemes.
\end{abstract}

\section{Introduction}
The rapid development and widespread adoption of large language models (LLMs) have made high-quality machine-generated text increasingly accessible, while also raising concerns about misuse, including misinformation, spam, and academic misconduct (Weidinger et al., 2022; Mitchell et al., 2023; Wang et al., 2025b; Liang et al., 2026; Lalai et al., 2025). Reliable identification of machine-generated content has therefore become an increasingly important problem (Wang et al., 2025b; Liang et al., 2026). Watermarking offers a proactive solution by embedding hidden statistical signals during generation that can later be verified by a detector (Christ et al., 2024; Kuditipudi et al., 2024; Zhao et al., 2024; Mao et al., 2025; Chen et al., 2025; Huang \& Wan, 2025).

A substantial number of methods have been proposed for the watermark detection problem, with diverse underlying intuitions. The mainstream families include logit-based methods (Kirchenbauer et al., 2023; Nemecek et al., 2026; Si et al., 2026), sampling-based methods (Hu et al., 2024; Kuditipudi et al., 2024; Mao et al., 2025; Chen et al., 2025; Huang \& Wan, 2025), and semantic or transformation-based methods (Liu et al., 2024; Dabiriaghdam \& Wang, 2025; Ye et al., 2026; Park et al., 2026). Despite these differences in mechanism, most of these approaches are designed to serve the same goal: document-level classification, i.e., determining whether a document is watermarked as a whole (Lalai et al., 2025; Liang et al., 2026; Bashir, 2026).

Real documents are often mixtures of human- and machine-generated content (Pan et al., 2025; Zhao et al., 2025; Si et al., 2026; Bashir, 2026). For example, a policy report or public-safety document may contain a small number of LLM-generated paragraphs that contain misinformation or fake instructions (Wang et al., 2025b; Liang et al., 2026). A document-level detector may reveal that watermarked content exists somewhere in the document, but fails to identify which parts are potentially machine-generated (Zhao et al., 2025; Si et al., 2026; Bashir, 2026). In such high-stakes settings, reliable localization is essential for targeting the specific watermarked paragraphs. Currently, watermark localization still remains relatively unexplored, with only a few existing works (Pan et al., 2025; Zhao et al., 2025; Si et al., 2026). WinMax (Kirchenbauer et al., 2024) first scores the sequence on a per-token basis and then searches over contiguous windows for the strongest watermark statistic. WaterSeeker (Pan et al., 2025) uses a coarse-to-fine strategy, first extracting a document-wide anomaly signal to identify suspicious regions and then refining them through local traversal. These methods share a common reliance on dense token-level watermark scoring across the entire document. Such document-wide scoring can become costly when obtaining token-level watermark evidence itself involves computationally intensive detectors (Block et al., 2025; Huang \& Wan, 2025). Then a question naturally arises:
\begin{center}\itshape
How can we significantly reduce the query cost while maintaining good detection performance?
\end{center}

To address this problem, this paper starts from a simple observation: locating a watermark does not necessarily require inspecting every token. Watermarked content usually occupies only part of a mixed-source document (Pan et al., 2025; Zhao et al., 2025; Si et al., 2026), and the corresponding detection signals tend to be concentrated around those regions (Pan et al., 2025; Zhao et al., 2025). Therefore, instead of scoring the entire document densely, we ask whether a small number of carefully chosen local observations can be used to recover the overall watermark localization. Intuitively, we first inspect the document broadly, then devote more of the remaining budget to regions that show stronger watermark evidence and to their nearby regions (Pan et al., 2025; Zhao et al., 2025; Locatelli et al., 2016). In this way, the detector gradually forms a global picture of where the watermark is likely to appear from only a fraction of the local observations.

To instantiate the adaptive querying strategy above without tying it to a particular watermark family, we map the detection signal at each queried token to a local e-value (Vovk \& Wang, 2021), with larger values indicating stronger evidence for the existence of a watermark (Shafer et al., 2011; Vovk \& Wang, 2021; Ramdas et al., 2023). Unlike previous works, to locate watermarks, we adopt a statistical hypothesis testing framework based on the e-value. This framework fits our setting for two reasons: it provides a watermark-family-invariant representation of evidence and, at the same time, supports valid evidence accumulation under adaptive querying (Shafer et al., 2011; Ramdas et al., 2023; Gr\"unwald et al., 2024). The queried local e-values within each region are then combined into a region-level e-value that summarizes the region's whole watermark evidence (Vovk \& Wang, 2021; Gr\"unwald et al., 2024). Once the query budget is exhausted, we must translate the accumulated regional evidence into a set of localized watermark segments. This naturally gives rise to a multiple hypothesis testing problem over candidate regions. Note that the sizes of watermark regions are heterogeneous. Inspired by Wang \& Ramdas (2022), we propose a novel token-weighted e-BH procedure to address this problem. Besides, to reduce the querying cost, we introduce a new strategy for budget allocation. Combining the region-level e-value, token-weighted e-BH procedure, and new budget allocation scheme, the resulting framework is called BELL (Budgeted E-evidence Localization for LLM Watermarks).

Beyond query efficiency, BELL enjoys a finite-sample statistical guarantee: it controls the token-weighted false discovery rate (FDR) at a prescribed level, without independence assumptions across blocks, tokens, or query rounds (Wang \& Ramdas, 2022). This guarantee is particularly important for localization, as it controls the proportion of human-written tokens that are falsely identified as watermarked among all reported watermark regions (Benjamini \& Hochberg, 1995). Extensive experiments across multiple LLMs, datasets, and watermark families further demonstrate the effectiveness of BELL. For instance, for the WikiText-103 dataset watermarked by Mistral-7B through the Hu Unbiased method, to reach a target recall\footnote{In this paper, the watermarked tokens or blocks are considered to be positive.} of 90\% while satisfying the FDR constraint, WaterSeeker (Pan et al., 2025) needs around 90\% of the total tokens to be queried, while BELL needs only 23\% of the document tokens to be queried, which is a 67\% reduction in queries. Our main contributions are summarized as follows:
\begin{itemize}
\item \textbf{Budgeted localization.} To the best of our knowledge, we are the first to explore reducing computational cost in LLM watermark localization from the perspective of a limited query budget.
\item \textbf{General framework with statistical guarantees.} To reduce the query cost, we propose a novel BELL framework with adaptive budget allocation, which can control token-weighted FDR for arbitrarily dependent block-level e-values.
\item \textbf{Empirical effectiveness.} Extensive experimental results demonstrate the superiority of our method over existing algorithms across multiple LLMs, datasets, and watermark families.
\end{itemize}

\section{Preliminaries}
\para{Notations.} Consider a document of $N$ tokens $x_{1:N}$, partitioned into $m$ ordered contiguous blocks $I_1,\ldots,I_m$ and $[m]=\{1,\ldots,m\}$. For each $j\in[m]$, block $I_j$ has length $L_j=|I_j|$ and $\sum_{j=1}^m L_j=N$. Let $\theta_j\in[0,1]$ denote the watermark intensity of block $I_j$, i.e., the fraction of tokens in $I_j$ originating from watermarked generation. Larger $\theta_j$ indicates a greater presence of watermarked content. Given a pre-specified threshold $\theta_0$, blocks with $\theta_j\leq\theta_0$ are treated as unwatermarked blocks, while blocks with $\theta_j>\theta_0$ are watermarked blocks. We denote
\[
\Hnull=\{j:\theta_j\leq\theta_0\},\qquad \Halt=\{j:\theta_j>\theta_0\}.
\]
$\Hnull$ and $\Halt$ represent the sets of indices of the unwatermarked blocks and watermarked blocks, respectively. Our goal is to identify the token regions in $\Halt$ effectively with less computational cost. We then consider a budgeted setting in which the verifier can evaluate the local detector for at most $B\leq N$ token positions.

\para{Formulation of the problem.} As discussed in Section 1, we investigate watermark localization from the perspective of hypothesis testing. Specifically, for $j\in[m]$, the watermark localization task is formulated as the following multiple hypothesis testing problem:
\begin{equation}\label{eq:hypotheses}
\begin{aligned}
H_{j;0}&:\text{ Block $j$ is unwatermarked }(\theta_j\leq\theta_0),\\
H_{j;1}&:\text{ Block $j$ is watermarked }(\theta_j>\theta_0)
\end{aligned}
\end{equation}
where $H_{j,0}$ and $H_{j,1}$ are called the null hypothesis and alternative hypothesis, respectively. If $H_{j,0}$ is rejected, we declare that block $I_j$ is watermarked. At round $t$, the verifier chooses a block $I_{j_t}$ and an unqueried token $x_{a_t}\in I_{j_t}$. The verifier may make the choice based on all the previously observed evidence. But the choice is irrevocable and no offline information is available to the verifier.

Querying $x_{a_t}$ reveals a watermark-specific statistical signal $Z_{x_{a_t}}$. Since different watermark families produce different local statistics (Kirchenbauer et al., 2023; Hu et al., 2024; Dathathri et al., 2024; Kuditipudi et al., 2024), we seek a common statistical representation of evidence to solve the problem without being tied to a certain watermark family. In statistics, p-values are often used to make decisions for problems in Eq.\ (1). However, ordinary p-values are less convenient in our setting because evidence is acquired sequentially and the ground-truth distribution of data is unknown (Shafer et al., 2011; Howard et al., 2021; Ramdas et al., 2023; Gr\"unwald et al., 2024). We therefore use e-values (Shafer et al., 2011; Vovk \& Wang, 2021; Wang \& Ramdas, 2022; Ramdas et al., 2023; Gr\"unwald et al., 2024), which can adaptively accumulate the localization signal under conditional local e-validity, without requiring independence across query rounds. The definition of an e-value is as follows.
\begin{definition}[e-value (Vovk \& Wang, 2021)]
Given a sample $\widetilde X$, a nonnegative statistic $e(\widetilde X)$ is called an e-value under the null hypothesis $H_0$ if $\E[e(\widetilde X)\mid H_0]\leq1$.
\end{definition}

Contrary to p-values, larger e-values indicate stronger evidence against the null hypothesis. At each queried token $x_{a_t}$, the verifier receives the detection signal $Z_{x_{a_t}}$. We need to construct the watermark-specific adapter $\Aw$ to convert the detector signal $Z_{x_{a_t}}$ into a local e-value $e_{x_{a_t}}=\Aw(Z_{x_{a_t}})$.

The likelihood-ratio (LR) adapter and moment-generating-function (MGF)-based adapter are two useful strategies to construct e-values. The LR adapter applies when both the null and watermarked likelihoods are available (Shafer et al., 2011; Vovk \& Wang, 2021), whereas the MGF-based adapter is more suitable for watermark families whose null behavior can be characterized without specifying a full alternative distribution (Vovk \& Wang, 2021; Gr\"unwald et al., 2024). For the widespread detection signals used in Hu et al.\ (2024), Dathathri et al.\ (2024), and Kuditipudi et al.\ (2024), we design appropriate adapters to convert them to e-values. Detailed discussions are deferred to Appendix D.4.

Once the query budget $B$ is exhausted, we then aggregate the local signals within each block to obtain a block-level detection statistic by taking their product. Denote by $\mathcal T_j(B):=\{t\in[B]:j_t=j\}$ the set of query rounds allocated to block $I_j$. Formally, the block-level e-value can be expressed as $E_j(B)=\prod_{t\in\mathcal T_j(B)}e_t$, where $E_j(0)=1$. It is easy to verify that $E_j(B)=\prod_{t\in\mathcal T_j(B)}e_t$ is still a valid e-value.

The e-BH procedure (Wang \& Ramdas, 2022) is a popular method for solving multiple hypothesis testing problems. Its definition is as follows.
\begin{definition}[e-BH Procedure (Wang \& Ramdas, 2022)]
Given the e-values $E_1,E_2,\ldots,E_m$ corresponding to the null hypotheses $H_{1,0},H_{2,0},\ldots,H_{m,0}$, let $E_{(i)}$ be the $i$-th order statistic from the largest to the smallest. For a pre-specified level $\alpha\in(0,1)$, define
\[
i^*_{\mathrm{eBH}}=\max\left\{i\in[m]:E_{(i)}(B)\geq\frac{m}{\alpha i}\right\},
\]
with $i^*_{\mathrm{eBH}}=0$ if the set is empty. Then, the null hypothesis $H_{(i),0}$ is rejected if $i\leq i^*_{\mathrm{eBH}}$.
\end{definition}

The e-BH procedure can control the false discovery rate (FDR) at the prescribed level $\alpha$ (Benjamini \& Hochberg, 1995; Wang \& Ramdas, 2022). In practice, $\alpha$ is often set to 0.05. The FDR is a standard error criterion in the multiple hypothesis testing literature (Wang \& Ramdas, 2022; Xu \& Ramdas, 2024; Zhang et al., 2025; Wang et al., 2025a; Sun \& Wang, 2026). Let $\Shat\subseteq[m]$ denote the index set of rejected hypotheses. The FDR is defined as the expectation of the false discovery proportion (FDP), namely $\FDR=\E[\FDP]$, where $\FDP=\frac{|\Shat\cap\Hnull|}{|\Shat|\vee1}$. FDR measures the expected proportion of false discoveries among all rejected hypotheses. In watermark localization, controlling FDR limits spurious watermark detections and prevents unwatermarked regions from being excessively reported as watermarked. Although evidence collected from different blocks may be statistically dependent, e-BH can still provide FDR control based on block-level e-values.

\section{Token-weighted FDR and e-BH Procedure}
Standard FDR assigns equal importance to every rejected block regardless of its length. However, this is not fully aligned with watermark localization: falsely declaring a longer null block as watermarked is more serious than making the same error on a shorter null block. Motivated by weighting ideas in Benjamini \& Hochberg (1997) and Genovese et al.\ (2006), we propose the novel token-weighted FDR and token-weighted e-BH procedure to address the issue of varying lengths.
\begin{definition}[Token-weighted FDR]
For each block $I_j$, let $L_j=|I_j|$ denote its number of tokens. Given the index set of rejected blocks $\Shat$, we define
\begin{equation}\label{eq:fdr}
\FDP_L=\frac{\sum_{j\in\Shat\cap\Hnull}L_j}{\left(\sum_{j\in\Shat}L_j\right)\vee1},\qquad
\FDR_L=\E[\FDP_L].
\end{equation}
\end{definition}
Unlike standard FDR, $\FDR_L$ becomes length-aware and gives more weight to the blocks with a higher number of tokens, making the error criterion better aligned with the token-level nature of watermark localization.

To control $\FDR_L$, we generalize the e-BH procedure to accommodate heterogeneous block lengths. Let $L_{(k)}$ denote the length of the block associated with the $k$-th largest e-value $E_{(k)}(B)$, and define the cumulative token mass of the top-$k$ blocks as $C_k=\sum_{\ell=1}^k L_{(\ell)}$.
\begin{definition}[Token-weighted e-BH Procedure]
Given the block-level e-values $E_1(B),\allowbreak E_2(B),\allowbreak\ldots,\allowbreak E_m(B)$ corresponding to the null hypotheses $H_{1,0},H_{2,0},\ldots,H_{m,0}$, let $E_{(i)}(B)$ be the $i$-th order statistic from the largest to the smallest. For a pre-specified level $\alpha\in(0,1)$ and total number of tokens $N=\sum_{j=1}^m L_j$, define
\[
i^*_{\mathrm{TEBH}}=\max\left\{i\in[m]:E_{(i)}(B)\geq\frac{N}{\alpha C_i}\right\},
\]
with $i^*_{\mathrm{TEBH}}=0$ if the set is empty. Then, the null hypothesis $H_{(i),0}$ is rejected if $i\leq i^*_{\mathrm{TEBH}}$.
\end{definition}
Unlike standard e-BH, the decision threshold of token-weighted e-BH depends on both the cumulative token mass $C_k$ and the number of selected blocks. This allows our proposed procedure to account for heterogeneous block lengths and align the rejection rule with the token-weighted FDR. The following theorem proves that token-weighted e-BH can control $\FDR_L$ at a prescribed level $\alpha\in(0,1)$ without the assumption of independence.
\begin{theorem}[FDR Control of Token-weighted e-BH]\label{thm:fdr}
Given arbitrarily dependent block-level e-values $E_1(B),E_2(B),\ldots,E_m(B)$ corresponding to the null hypotheses $H_{1,0},H_{2,0},\ldots,H_{m,0}$, for a pre-specified level $\alpha\in(0,1)$, the token-weighted e-BH procedure satisfies
\[
\FDR_L=\E\left[\frac{\sum_{j\in\Shat\cap\Hnull}L_j}{(\sum_{j\in\Shat}L_j)\vee1}\right]\leq\alpha.
\]
\end{theorem}
The proof of Theorem 3.3 is presented in Appendix C.2. Theorem 3.3 provides a statistical safeguard against spurious token-level watermark detection. With this error-control guarantee, the query allocation scheme can focus on improving localization power and query efficiency. Different allocation strategies may change where the limited query budget is spent, but they do not affect the token-weighted FDR control.

\section{Allocation Scheme}
The remaining key issue is how to allocate a fixed budget $B\leq N$ across the document. Given a fixed query budget, our allocation follows a broad-to-focused strategy: we first maintain coarse coverage of the document, and then progressively concentrate queries on blocks with stronger observed evidence or watermarked neighbors, or blocks close to the final decision threshold. Existing adaptive allocation methods cannot be directly plugged in because they typically allocate based on each block's own reward--uncertainty tradeoff, whereas here each block has a limited query capacity, and its allocation priority also depends on neighboring evidence and proximity to the final testing threshold.

Each token can be queried at most once. Let $n_j(t)=\sum_{s<t}\1\{J_s=j\}$ denote the number of tokens already queried from block $j$ before round $t$, so that $L_j-n_j(t)$ positions remain available. A block may nevertheless be selected repeatedly, with each visit querying a previously unseen token. Define
\begin{equation}
Y_s=\clip(\log(e_s\vee\epsilon),-C_-,C_+),\qquad
\widehat\mu_j(t)=\frac{\sum_{s<t:J_s=j}Y_s}{n_j(t)\vee1}.
\end{equation}
Here $\clip(x,a,b)=\min\{\max\{x,a\},b\}$, $\epsilon>0$ avoids $\log0$, and $C_-,C_+>0$ bound extreme log-evidence. These clipped statistics are used only for allocation; terminal testing always uses the exact products $E_j(B)$. Our scheme combines four mechanisms: coverage, optimistic exploration, neighborhood expansion, and near-confirmation.

\para{Coverage.} Because the block evidence $E_j(B)$ in Section 2 is accumulated only from queried tokens, concentrating too early may leave some blocks with insufficient evidence. We therefore first establish broad coverage (Locatelli et al., 2016; Pan et al., 2025). Each block receives an initial target
\begin{equation}
b_j=\min\left\{L_j,\max\{b_{\min},\lceil c_{\mathrm{cov}}L_j^\beta\rceil\}\right\},
\end{equation}
where $b_{\min}$ sets the minimum number of initial queries, $c_{\mathrm{cov}}$ controls the overall scale, and $\beta$ controls how the target grows with block length. Until these targets are satisfied, the scheme prioritizes the block with the smallest normalized coverage $n_j(t)/(b_j\vee1)$. Afterwards, at the fixed coverage rounds $\mathcal T_{\mathrm{cov}}\subseteq[B]$, the least-covered non-exhausted block is selected according to $n_j(t)/L_j^\beta$, reducing persistent blind spots caused by noisy early evidence.

\para{Optimistic exploration.} After broad coverage, we allocate according to the sequential evidence defined in Section 2. Since $\widehat\mu_j(t)$ is the empirical mean of the clipped log-evidence observed in block $I_j$, we use the UCB-style score (Auer et al., 2002; Locatelli et al., 2016)
\begin{equation}
U_j(t)=\widehat\mu_j(t)+(C_-+C_+)\sqrt{\frac{\log(4mt^2/\delta)}{2(n_j(t)\vee1)}},
\end{equation}
where $\delta\in(0,1)$ controls the exploration bonus. The empirical term favors blocks with stronger observed evidence, while the bonus is larger for under-sampled blocks and prevents uncertain regions from being discarded too early.

\para{Neighborhood expansion.} Section 2 models the document as ordered contiguous blocks, while watermark signals often extend across adjacent blocks (Kirchenbauer et al., 2024; Pan et al., 2025; Zhao et al., 2025). We therefore use neighboring evidence to guide where to query next. With $\mathcal N(j)=\{j-1,j+1\}\cap[m]$ and $[x]_+=\max\{x,0\}$, define
\begin{equation}
V_j(t)=\frac{\sum_{\ell\in\mathcal N(j)}\rho_\ell(t)[\widehat\mu_\ell(t)]_+}{\left(\sum_{\ell\in\mathcal N(j)}\rho_\ell(t)\right)\vee1},\qquad
\rho_\ell(t)=\min\left\{1,\frac{n_\ell(t)}{b_\ell\vee1}\right\}.
\end{equation}
Here $\rho_\ell(t)$ is a coverage weight, so evidence from a sparsely queried neighbor contributes less. The allocation score is $A_j(t)=U_j(t)+\eta V_j(t)$, where $\eta\geq0$ controls neighborhood influence. Neighbor evidence affects allocation only and is never transferred into another block's test statistic.

\para{Near-confirmation.} Section 4 determines discoveries through the token-weighted e-BH threshold. Once promising blocks emerge, we therefore favor blocks for which a small number of additional queries can plausibly change the terminal decision (Wang \& Ramdas, 2022; Locatelli et al., 2016). Let $C_{\mathrm{plan}}=\max\{1,\lceil\kappa N\rceil\}$ be a planning value for the eventual rejected token mass, where $\kappa\in(0,1]$ specifies the assumed fraction of document tokens that will eventually be rejected, and define $\tau_{\mathrm{plan}}=\log\bigl(N/(\alpha C_{\mathrm{plan}})\bigr)$. We estimate
\begin{equation}
\widehat h_j(t)=\frac{[\tau_{\mathrm{plan}}-\log E_j(t-1)]_+}{\max\{U_j(t),\varepsilon\}}.
\end{equation}
Thus, $\widehat h_j(t)$ estimates how many additional queries are needed to reach the planning threshold if future log-evidence accumulates at the optimistic rate $U_j(t)$. Block $I_j$ is near-confirmable if $U_j(t)>0$, $\widehat h_j(t)\leq H$, and $\widehat h_j(t)\leq L_j-n_j(t)$, where $H$ is the confirmation horizon and the last condition enforces the remaining query capacity. When such blocks exist, refinement is restricted to them; otherwise the policy maximizes $A_j(t)$ over all non-exhausted blocks.

Combining the block-level e-values, token-weighted e-BH procedure, and the proposed allocation scheme, our practical localization method, called BELL, is presented in Algorithm 1. Overall, the proposed algorithm has a strict theoretical guarantee of $\FDR_L$ control without additional strong assumptions, and extensive experiments show its superiority in the next section.

\begin{algorithm}[tbp]
\caption{BELL}\label{alg:bell}
\begin{algorithmic}[1]
\Require Blocks $\{I_j,L_j\}_{j=1}^m$, budget $B$, adapter $\Aw$, level $\alpha$
\State Compute coverage targets $b_j$; initialize $n_j\gets0$ and $E_j\gets1$
\For{$t=1,\ldots,B$}
\State $\mathcal B_t\gets\{j:n_j<L_j\}$
\If{some $j\in\mathcal B_t$ satisfies $n_j<b_j$}
\State $J_t\in\argmin_{j\in\mathcal B_t:n_j<b_j}\frac{n_j}{b_j\vee1}$
\ElsIf{$t\in\mathcal T_{\mathrm{cov}}$}
\State $J_t\in\argmin_{j\in\mathcal B_t}\frac{n_j}{L_j^\beta}$
\Else
\State Compute $U_j(t)$, $V_j(t)$, $A_j(t)$, and $\widehat h_j(t)$ for $j\in\mathcal B_t$
\State $\mathcal C_t\gets\{j\in\mathcal B_t:j\text{ is near-confirmable}\}$
\If{$\mathcal C_t\ne\varnothing$}
\State $J_t\in\argmax_{j\in\mathcal C_t}A_j(t)$
\Else
\State $J_t\in\argmax_{j\in\mathcal B_t}A_j(t)$
\EndIf
\EndIf
\State Query an unqueried token $a_t\in I_{J_t}$ and obtain $e_t=\Aw(Z_t)$
\State $E_{J_t}\gets E_{J_t}e_t$; $n_{J_t}\gets n_{J_t}+1$; update allocation statistics
\EndFor
\State Apply token-weighted e-BH to $\{E_j(B),L_j\}_{j=1}^m$ and obtain the rejection set $\Shat$
\State \textbf{Output:} Declare $I_j$ watermarked (reject $H_{j,0}$) if $j\in\Shat$, and the rest unwatermarked.
\end{algorithmic}
\end{algorithm}

\section{Experiments}
\subsection{Experimental Setup}
\para{Models, datasets, and documents.} We use Qwen2.5-7B-Instruct (Yang et al., 2024) and Mistral-7B-Instruct-v0.3 to generate watermarked content (Jiang et al., 2023), with source text from C4 (Raffel et al., 2020) and WikiText-103 (Merity et al., 2016). Each 4096-token document is partitioned into $m=40$ variable-length blocks and contains five spatially separated, non-overlapping generated regions. Each region contains 160--240 tokens and the five region-level watermark intensities are a random permutation of $\{1.0,1.0,0.75,0.75,0.5\}$. Neighboring generated regions are separated by at least 128 source tokens, and their boundaries are sampled independently of the block partition. We use $\theta_0=0.6$ and full construction details are given in Appendix D.1.

\para{Watermarking schemes.} We evaluate four token-level watermark families: KGW (Kirchenbauer et al., 2023), Hu Unbiased (Hu et al., 2024), SynthID-style (Dathathri et al., 2024), and Gumbel (Kuditipudi et al., 2024). Each detector is mapped to local e-evidence following the definitions in Section 2, and we detail the constructions in Appendix D.4.

\para{Baselines.} We fairly adapt existing watermark localization methods to the budgeted setting, and the details are deferred to Appendix D.3. B-WinMax, based on WinMax (Kirchenbauer et al., 2024), sparsely probes the document and refines high-score windows; B-WaterSeeker, adapted from WaterSeeker (Pan et al., 2025), uses sparse screening followed by local refinement; and B-AOL, adapted from the adaptive online locator of Zhao et al.\ (2025), adaptively focuses queries on promising regions to refine segment locations. For convenience, we still refer to them as WinMax, WaterSeeker, and AOL in the rest of the paper.

\para{Metrics.} We evaluate token-level precision, recall, $F_1$, and intersection-over-union (IoU). We define the query fraction as $B/N$, which is the ratio of the budget $B$ to the total length $N$. For a target recall $\rho$, we additionally introduce a metric as follows:
\begin{equation}
B^\star(\rho)=\min\{B/N:\Recall_{\mathrm{tok}}(B)\geq\rho,\ \FDR_L(B)\leq\alpha\},
\end{equation}
which quantifies the minimum query cost needed to attain a target recall $\rho$ under FDR control.

\subsection{Comparison with Baselines}
\para{Query efficiency.} Figure 1 reports the minimum query fraction $B/N$ required to achieve different target recall levels on Qwen2.5-7B-Instruct with C4. As shown in Figure 1, compared with the competing methods WaterSeeker, WinMax, and AOL, our proposed BELL achieves the same recall levels with substantially lower query costs across KGW, Hu Unbiased, SynthID-style, and Gumbel. For example, at 80\% recall, BELL requires a query fraction of only 12\%--14\%, compared with 38\%--45\% for WinMax, 42\%--50\% for WaterSeeker, and 44\%--52\% for AOL. Moreover, the advantage of BELL becomes more pronounced at higher recall levels. At 90\% recall, our method requires a query fraction of only 21\%--23\%, whereas the competing methods require 80\%--89\%. Additional experimental results in Figures 3, 4, and 5 of Appendix E.1 exhibit the same trends as those shown in Figure 1. These results demonstrate that our proposed BELL consistently requires a smaller query budget to attain high recall across different watermark families and LLMs.

\figpaper{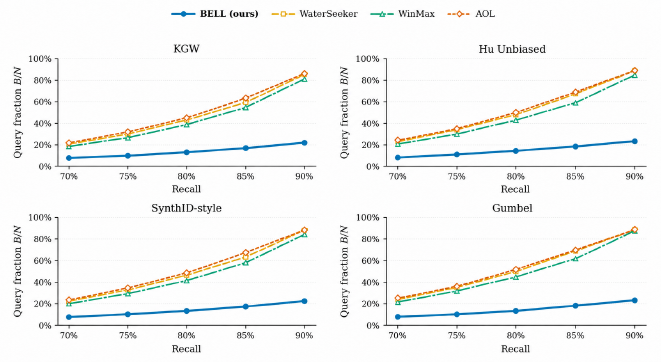}{0.90}{Comparison of the minimum query fraction $B/N$ required to achieve different target recall levels across various algorithms. All results are evaluated using Qwen2.5-7B-Instruct on C4.}{fig:1}
\begin{table}[htbp]
\centering
\caption{Comparison of different methods at a fixed query fraction of $B/N=20\%$ with various watermarking schemes. All results are evaluated using Qwen2.5-7B-Instruct on C4.}
\label{tab:1}
\vspace{4pt}
\resizebox{\linewidth}{!}{%
\begin{tabular}{@{}l*{10}{c}@{}}
\toprule
\multirow{2}{*}{Method} & \multicolumn{5}{c}{KGW} & \multicolumn{5}{c}{Hu Unbiased} \\
 & Prec.\,$\uparrow$ & Rec.\,$\uparrow$ & F1\,$\uparrow$ & IoU\,$\uparrow$ & $B^{\ast}(0.8)\,\downarrow$ & Prec.\,$\uparrow$ & Rec.\,$\uparrow$ & F1\,$\uparrow$ & IoU\,$\uparrow$ & $B^{\ast}(0.8)\,\downarrow$ \\
\midrule
WaterSeeker & 84.2 & 69.7 & 76.3 & 61.6 & 42 & 82.3 & 61.8 & 70.6 & 54.6 & 48 \\
WinMax & 86.1 & 72.0 & 78.4 & 64.5 & 38 & 86.0 & 66.7 & 75.1 & 60.2 & 43 \\
AOL & 84.3 & 67.4 & 74.9 & 59.9 & 44 & 84.3 & 57.7 & 68.5 & 52.1 & 50 \\
\midrule
\textbf{BELL (ours)} & \textbf{95.4} & \textbf{89.9} & \textbf{92.6} & \textbf{86.2} & \textbf{12} & \textbf{94.4} & \textbf{87.9} & \textbf{91.0} & \textbf{83.6} & \textbf{14} \\
\bottomrule
\addlinespace[7pt]
\toprule
\multirow{2}{*}{Method} & \multicolumn{5}{c}{SynthID-style} & \multicolumn{5}{c}{Gumbel} \\
 & Prec.\,$\uparrow$ & Rec.\,$\uparrow$ & F1\,$\uparrow$ & IoU\,$\uparrow$ & $B^{\ast}(0.8)\,\downarrow$ & Prec.\,$\uparrow$ & Rec.\,$\uparrow$ & F1\,$\uparrow$ & IoU\,$\uparrow$ & $B^{\ast}(0.8)\,\downarrow$ \\
\midrule
WaterSeeker & 83.8 & 63.4 & 72.2 & 56.5 & 46 & 82.3 & 58.9 & 68.7 & 52.3 & 50 \\
WinMax & 86.2 & 69.6 & 77.0 & 62.6 & 41 & 85.5 & 64.7 & 73.7 & 58.3 & 45 \\
AOL & 85.4 & 60.7 & 71.0 & 55.0 & 48 & 83.8 & 56.8 & 67.7 & 51.2 & 52 \\
\midrule
\textbf{BELL (ours)} & \textbf{94.8} & \textbf{88.9} & \textbf{91.8} & \textbf{84.8} & \textbf{13} & \textbf{94.6} & \textbf{87.9} & \textbf{91.1} & \textbf{83.7} & \textbf{13} \\
\bottomrule
\end{tabular}
}
\end{table}

To comprehensively evaluate the effectiveness of the proposed method, Table 1 reports localization performance across different metrics at a fixed query fraction of $B/N=20\%$. From Table 1, we observe that BELL consistently outperforms the competing methods across all four watermarking schemes. Specifically, compared with AOL under the Gumbel setting, our BELL improves the F1 score from 67.7\% to 91.1\%, with a \textbf{23.4\%} improvement. Averaged over the four watermarking schemes, BELL obtains 94.8\% precision, 88.7\% recall, 91.6\% F1, and 84.6\% IoU, improving over the strongest baseline, WinMax, by \textbf{8.9\%}, \textbf{20.4\%}, \textbf{15.6\%}, and \textbf{23.2\%}, respectively. The consistent improvements observed across KGW, Hu Unbiased, SynthID-style, and Gumbel indicate that BELL generalizes effectively across watermark families with distinct underlying detection characteristics. The experimental results evaluated based on Mistral-7B-Instruct-v0.3 and other datasets are presented in Tables 5, 6, and 7 in Appendix E.3. These results further support our conclusion. Due to space limitations, the experimental results on FDR are deferred to Table 4 of Appendix E.2. As shown in Table 4, the empirical FDR of BELL remains below the prescribed significance level $\alpha=0.05$ for different LLMs, watermark families, datasets, and query fractions $B/N$, providing empirical support for the FDR control guarantee established in Theorem 3.3.

\subsection{Why Does BELL Work?}
In this subsection, we explore the underlying reason why BELL has superior performance. To quantitatively evaluate query efficiency, we introduce two metrics: the target query ratio $\TQR(B)=B^{-1}\sum_{t=1}^B\1\{J_t\in S^\star\}$ and the evidence gain $\Gain(B)=B^{-1}\sum_{t=1}^B[\log e_t]_+$. $\TQR(B)$ measures the fraction of the query budget allocated to ground-truth watermarked regions, and thus reflects how effectively the allocation policy concentrates queries on relevant blocks. $\Gain(B)$ measures the average positive log-evidence obtained per query, characterizing how much useful watermark evidence is extracted from each detector evaluation. Together, these metrics characterize whether an allocation strategy queries the right regions and whether those queries yield informative evidence. The corresponding experimental results are presented in Figure 2. From Figure 2, we observe that under the same query budget, BELL is far more efficient than the baselines across different watermark families. The same trend is also shown in Figures 6, 7, and 8 of Appendix E.4, which further shows our algorithm's robustness.
\figpaper{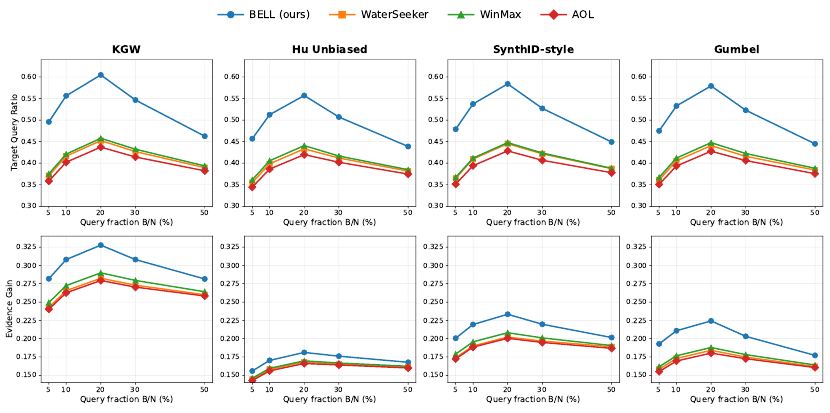}{1}{Evaluation of different methods' query efficiency through the target query ratio (TQR) and evidence gain (EG). All results are evaluated using Qwen2.5-7B-Instruct on C4.}{fig:2}
\begin{table}[htbp]
\centering
\caption{Ablation results (\%) of the query-budget allocation scheme on Qwen2.5-7B-Instruct with C4.}
\label{tab:2}
\vspace{4pt}
\resizebox{\linewidth}{!}{%
\begin{tabular}{@{}l*{12}{c}@{}}
\toprule
\multirow{2}{*}{Variant} & \multicolumn{3}{c}{KGW} & \multicolumn{3}{c}{Hu Unbiased} & \multicolumn{3}{c}{SynthID-style} & \multicolumn{3}{c}{Gumbel} \\
\cmidrule(lr){2-4}\cmidrule(lr){5-7}\cmidrule(lr){8-10}\cmidrule(lr){11-13}
 & Recall & IoU & EG & Recall & IoU & EG & Recall & IoU & EG & Recall & IoU & EG \\
\midrule
w/o Coverage & 80.9 & 77.5 & 28.9 & 77.6 & 73.8 & 16.0 & 79.2 & 75.6 & 20.6 & 79.2 & 75.5 & 19.8 \\
w/o Optimism & 84.7 & 81.1 & 29.8 & 82.1 & 78.1 & 16.5 & 83.4 & 79.5 & 21.2 & 82.9 & 79.0 & 20.4 \\
w/o Neighbor & 83.4 & 79.9 & 30.3 & 80.9 & 77.0 & 16.8 & 82.1 & 78.4 & 21.6 & 81.9 & 78.0 & 20.7 \\
w/o Confirmation & 85.9 & 82.3 & 26.9 & 83.4 & 79.3 & 14.9 & 84.7 & 80.7 & 19.2 & 83.9 & 79.9 & 18.4 \\
\midrule
\textbf{Full BELL} & \textbf{89.9} & \textbf{86.2} & \textbf{32.8} & \textbf{87.9} & \textbf{83.6} & \textbf{18.1} & \textbf{88.9} & \textbf{84.8} & \textbf{23.3} & \textbf{87.9} & \textbf{83.7} & \textbf{22.4} \\
\bottomrule
\end{tabular}
}
\end{table}

\subsection{Ablation Study}
We study how query-budget allocation affects BELL by ablating coverage, optimism, neighborhood expansion, and near-confirmation at $B/N=0.2$. As shown in Table 2, removing coverage causes the largest drop in recall and IoU across all four watermark families, averaging 9.4\% and 9.0\%, respectively. Neighborhood expansion is the second most important component, highlighting the benefit of local continuity. Removing near-confirmation has a smaller impact on recall and IoU but causes the largest decrease in evidence gain, showing its role in query efficiency. Optimism provides a smaller but consistent improvement. Overall, these components play complementary roles. Similar trends hold for other watermark-dataset settings, with additional results in Appendix E.5.

\section{Conclusion}
This paper focuses on reducing computational cost in LLM watermark localization with a limited query budget. To tackle this problem, we propose a novel BELL algorithm from the perspective of statistical hypothesis testing. By adaptively allocating detector queries to promising regions and converting local statistical signals into e-values, BELL can control token-weighted FDR based on our proposed token-weighted e-BH procedure. Extensive experiments demonstrate that BELL consistently improves query efficiency and localization performance over existing methods.

\section*{AI Use Statement}
In this work, we used generative AI tools to polish the writing and to assist with coding. AI tools were not used to generate research ideas or to develop or verify theorems and proofs or to design experiments. We have reviewed all AI-assisted work and take responsibility for the final content of this paper, including any text, claims, code, figures, or other artifacts.

\section*{Reproducibility Statement}
We describe the information needed to reproduce the main results in Section 5 and Appendix D. The full code and dataset will be released on GitHub in the future.

\clearpage
\section*{References}
\begin{list}{}{\setlength{\leftmargin}{1em}\setlength{\itemindent}{-1em}\setlength{\itemsep}{5pt}}
\item \begin{samepage}
Peter Auer, Nicolò Cesa-Bianchi, and Paul Fischer. Finite-time analysis of the multiarmed bandit problem. Machine Learning, 47(2--3):235--256, 2002.
\par
\end{samepage}

\item \begin{samepage}
Shariq Bashir. Visual pattern-based watermarking for large language model generated text. Neurocomputing, 682:133429, 2026.
\par
\end{samepage}

\item \begin{samepage}
Yoav Benjamini and Yosef Hochberg. Controlling the false discovery rate: A practical and powerful approach to multiple testing. Journal of the Royal Statistical Society: Series B (Methodological), 57(1):289--300, 1995.
\par
\end{samepage}

\item \begin{samepage}
Yoav Benjamini and Yosef Hochberg. Multiple hypotheses testing with weights. Scandinavian Journal of Statistics, 24(3):407--418, 1997.
\par
\end{samepage}

\item \begin{samepage}
Yoav Benjamini and Daniel Yekutieli. The control of the false discovery rate in multiple testing under dependency. The Annals of Statistics, 29(4):1165--1188, 2001.
\par
\end{samepage}

\item \begin{samepage}
Adam Block, Alexander Rakhlin, and Ayush Sekhari. GaussMark: A practical approach for structural watermarking of language models. In Proceedings of the 42nd International Conference on Machine Learning, pp. 4507--4569, 2025.
\par
\end{samepage}

\item \begin{samepage}
Ruibo Chen, Yihan Wu, Junfeng Guo, and Heng Huang. Improved unbiased watermark for large language models. In Proceedings of the 63rd Annual Meeting of the Association for Computational Linguistics (Volume 1: Long Papers), pp. 20587--20601, 2025.
\par
\end{samepage}

\item \begin{samepage}
Miranda Christ, Sam Gunn, and Or Zamir. Undetectable watermarks for language models. In Proceedings of the Thirty-Seventh Conference on Learning Theory, pp. 1125--1139, 2024.
\par
\end{samepage}

\item \begin{samepage}
Amirhossein Dabiriaghdam and Lele Wang. SimMark: A robust sentence-level similarity-based watermarking algorithm for large language models. In Proceedings of the 2025 Conference on Empirical Methods in Natural Language Processing, pp. 30785--30806, 2025.
\par
\end{samepage}

\item \begin{samepage}
Sumanth Dathathri, Abigail See, Sumedh Ghaisas, Po-Sen Huang, Rob McAdam, Johannes Welbl, Vandana Bachani, Alex Kaskasoli, Robert Stanforth, Tatiana Matejovicova, Jamie Hayes, Nidhi Vyas, Majd Al Merey, Jonah Brown-Cohen, Rudy Bunel, Borja Balle, Taylan Cemgil, Zahra Ahmed, Kitty Stacpoole, Ilia Shumailov, Ciprian Baetu, Sven Gowal, Demis Hassabis, and Pushmeet Kohli. Scalable watermarking for identifying large language model outputs. Nature, 634 (8035):818--823, 2024.
\par
\end{samepage}

\item \begin{samepage}
Christopher R. Genovese, Kathryn Roeder, and Larry Wasserman. False discovery control with p-value weighting. Biometrika, 93(3):509--524, 2006.
\par
\end{samepage}

\item \begin{samepage}
Peter Grünwald, Rianne de Heide, and Wouter Koolen. Safe testing. Journal of the Royal Statistical Society: Series B (Statistical Methodology), 86(5):1091--1128, 2024.
\par
\end{samepage}

\item \begin{samepage}
Wassily Hoeffding. Probability inequalities for sums of bounded random variables. Journal of the American Statistical Association, 58(301):13--30, 1963.
\par
\end{samepage}

\item \begin{samepage}
Steven R. Howard, Aaditya Ramdas, Jon McAuliffe, and Jasjeet Sekhon. Time-uniform, nonparametric, nonasymptotic confidence sequences. The Annals of Statistics, 49(2):1055--1080, 2021.
\par
\end{samepage}

\item \begin{samepage}
Zhengmian Hu, Lichang Chen, Xidong Wu, Yihan Wu, Hongyang Zhang, and Heng Huang. Unbiased watermark for large language models. In The Twelfth International Conference on Learning Representations, 2024.
\par
\end{samepage}

\item \begin{samepage}
Baizhou Huang and Xiaojun Wan. WaterPool: A language model watermark mitigating trade-offs among imperceptibility, efficacy and robustness. In Proceedings of the 2025 Conference of the Nations of the Americas Chapter of the Association for Computational Linguistics: Human Language Technologies (Volume 1: Long Papers), pp. 4156--4182, 2025.
\par
\end{samepage}

\item \begin{samepage}
Albert Q. Jiang, Alexandre Sablayrolles, Arthur Mensch, Chris Bamford, Devendra Singh Chaplot, Diego de las Casas, Florian Bressand, Gianna Lengyel, Guillaume Lample, Lucile Saulnier, Lélio Renard Lavaud, Marie-Anne Lachaux, Pierre Stock, Teven Le Scao, Thibaut Lavril, Thomas Wang, Timothée Lacroix, and William El Sayed. Mistral 7b. arXiv preprint arXiv:2310.06825, 2023.
\par
\end{samepage}

\item \begin{samepage}
John Kirchenbauer, Jonas Geiping, Yuxin Wen, Jonathan Katz, Ian Miers, and Tom Goldstein. A watermark for large language models. In Proceedings of the 40th International Conference on Machine Learning, pp. 17061--17084, 2023.
\par
\end{samepage}

\item \begin{samepage}
John Kirchenbauer, Jonas Geiping, Yuxin Wen, Manli Shu, Khalid Saifullah, Kezhi Kong, Kasun Fernando, Aniruddha Saha, Micah Goldblum, and Tom Goldstein. On the reliability of watermarks for large language models. In The Twelfth International Conference on Learning Representations, 2024.
\par
\end{samepage}

\item \begin{samepage}
Rohith Kuditipudi, John Thickstun, Tatsunori Hashimoto, and Percy Liang. Robust distortion-free watermarks for language models. Transactions on Machine Learning Research, 2024.
\par
\end{samepage}

\item \begin{samepage}
Harsh Nishant Lalai, Aashish Anantha Ramakrishnan, Raj Sanjay Shah, and Dongwon Lee. From intentions to techniques: A comprehensive taxonomy and challenges in text watermarking for large language models. In Findings of the Association for Computational Linguistics: NAACL 2025, pp. 6162--6175, 2025.
\par
\end{samepage}

\item \begin{samepage}
Erich L. Lehmann and Joseph P. Romano. Testing Statistical Hypotheses. Springer, New York, 3rd edition, 2005. doi: 10.1007/0-387-27605-X.
\par
\end{samepage}

\item \begin{samepage}
Xiang Li, Feng Ruan, Huiyuan Wang, Qi Long, and Weijie J. Su. A statistical framework of watermarks for large language models: Pivot, detection efficiency and optimal rules. The Annals of Statistics, 53(1):322--351, 2025.
\par
\end{samepage}

\item \begin{samepage}
Yuqing Liang, Jiancheng Xiao, Wensheng Gan, and Philip S. Yu. Watermarking techniques for large language models: A survey. Artificial Intelligence Review, 59(2):74, 2026.
\par
\end{samepage}

\item \begin{samepage}
Aiwei Liu, Leyi Pan, Xuming Hu, Shiao Meng, and Lijie Wen. A semantic invariant robust watermark for large language models. In The Twelfth International Conference on Learning Representations, 2024.
\par
\end{samepage}

\item \begin{samepage}
Andrea Locatelli, Maurilio Gutzeit, and Alexandra Carpentier. An optimal algorithm for the thresholding bandit problem. In Proceedings of the 33rd International Conference on Machine Learning, pp. 1690--1698, 2016.
\par
\end{samepage}

\item \begin{samepage}
Minjia Mao, Dongjun Wei, Zeyu Chen, Xiao Fang, and Michael Chau. Watermarking large language models: An unbiased and low-risk method. In Proceedings of the 63rd Annual Meeting of the Association for Computational Linguistics (Volume 1: Long Papers), pp. 7939--7960, 2025.
\par
\end{samepage}

\item \begin{samepage}
Stephen Merity, Caiming Xiong, James Bradbury, and Richard Socher. Pointer sentinel mixture models. arXiv preprint arXiv:1609.07843, 2016.
\par
\end{samepage}

\item \begin{samepage}
Eric Mitchell, Yoonho Lee, Alexander Khazatsky, Christopher D. Manning, and Chelsea Finn. DetectGPT: Zero-shot machine-generated text detection using probability curvature. In Proceedings of the 40th International Conference on Machine Learning, pp. 24950--24962, 2023.
\par
\end{samepage}

\item \begin{samepage}
Alexander Nemecek, Yuzhou Jiang, and Erman Ayday. Topic-based watermarks for large language models. In Findings of the Association for Computational Linguistics: ACL 2026, pp. 24372--24402, 2026.
\par
\end{samepage}

\item \begin{samepage}
Jerzy Neyman and Egon S. Pearson. On the problem of the most efficient tests of statistical hypotheses. Philosophical Transactions of the Royal Society of London. Series A, Containing Papers of a Mathematical or Physical Character, 231:289--337, 1933.
\par
\end{samepage}

\item \begin{samepage}
Leyi Pan, Aiwei Liu, Yijian Lu, Zitian Gao, Yichen Di, Lijie Wen, Irwin King, and Philip S. Yu. WaterSeeker: Pioneering efficient detection of watermarked segments in large documents. In Findings of the Association for Computational Linguistics: NAACL 2025, pp. 2866--2882, 2025.
\par
\end{samepage}

\item \begin{samepage}
Shinwoo Park, Hyejin Park, Hyeseon An, and Yo-Sub Han. A linguistics-aware LLM watermarking via syntactic predictability. In Proceedings of the 64th Annual Meeting of the Association for Computational Linguistics (Volume 1: Long Papers), pp. 45629--45647, 2026.
\par
\end{samepage}

\item \begin{samepage}
Colin Raffel, Noam Shazeer, Adam Roberts, Katherine Lee, Sharan Narang, Michael Matena, Yanqi Zhou, Wei Li, and Peter J. Liu. Exploring the limits of transfer learning with a unified text-to-text transformer. Journal of Machine Learning Research, 21(140):1--67, 2020.
\par
\end{samepage}

\item \begin{samepage}
Aaditya Ramdas, Peter Grünwald, Vladimir Vovk, and Glenn Shafer. Game-theoretic statistics and safe anytime-valid inference. Statistical Science, 38(4):576--601, 2023.
\par
\end{samepage}

\item \begin{samepage}
Glenn Shafer, Alexander Shen, Nikolai Vereshchagin, and Vladimir Vovk. Test martingales, bayes factors and p-values. Statistical Science, 26(1):84--101, 2011.
\par
\end{samepage}

\item \begin{samepage}
Runnan Si, Xin Xie, Junxiao Wang, Xiulong Liu, Xiaoyi Tao, Xinyu Tong, Sheng Chen, Heng Qi, and Keqiu Li. Sequence-level watermarking for large language models. Knowledge-Based Systems, 347:116180, 2026.
\par
\end{samepage}

\item \begin{samepage}
Liulei Sun and Ruodu Wang. Admissibility and complete classes for false discovery rate control with e-values. arXiv preprint arXiv:2607.14380, 2026.
\par
\end{samepage}

\item \begin{samepage}
Vladimir Vovk and Ruodu Wang. e-values: Calibration, combination and applications. The Annals of Statistics, 49(3):1736--1754, 2021.
\par
\end{samepage}

\item \begin{samepage}
Hongjian Wang, Sanjit Dandapanthula, and Aaditya Ramdas. Anytime-valid FDR control with the stopped e-BH procedure. Statistics \& Probability Letters, 226:110512, 2025a. doi: 10.1016/j.spl.2025.110512.
\par
\end{samepage}

\item \begin{samepage}
Ruodu Wang and Aaditya Ramdas. False discovery rate control with e-values. Journal of the Royal Statistical Society: Series B (Statistical Methodology), 84(3):822--852, 2022.
\par
\end{samepage}

\item \begin{samepage}
Xuhong Wang, Haoyu Jiang, Yi Yu, Jingru Yu, Yilun Lin, Ping Yi, Yingchun Wang, Yu Qiao, Li Li, and Fei-Yue Wang. Building intelligence identification system via large language model watermarking: A survey and beyond. Artificial Intelligence Review, 58(8):249, 2025b.
\par
\end{samepage}

\item \begin{samepage}
Larry Wasserman, Aaditya Ramdas, and Sivaraman Balakrishnan. Universal inference. Proceedings of the National Academy of Sciences, 117(29):16880--16890, 2020.
\par
\end{samepage}

\item \begin{samepage}
Laura Weidinger, Jonathan Uesato, Maribeth Rauh, Conor Griffin, Po-Sen Huang, John Mellor, Amelia Glaese, Myra Cheng, Borja Balle, Atoosa Kasirzadeh, Courtney Biles, Sasha Brown, Zac Kenton, Will Hawkins, Tom Stepleton, Abeba Birhane, Lisa Anne Hendricks, Laura Rimell, William Isaac, Julia Haas, Sean Legassick, Geoffrey Irving, and Iason Gabriel. Taxonomy of risks posed by language models. In Proceedings of the 2022 ACM Conference on Fairness, Accountability, and Transparency, pp. 214--229, 2022.
\par
\end{samepage}

\item \begin{samepage}
Ziyu Xu and Aaditya Ramdas. Online multiple testing with e-values. In Proceedings of the 27th International Conference on Artificial Intelligence and Statistics, volume 238 of Proceedings of Machine Learning Research, pp. 3997--4005, 2024.
\par
\end{samepage}

\item \begin{samepage}
An Yang, Baosong Yang, Beichen Zhang, Binyuan Hui, Bo Zheng, Bowen Yu, Chengyuan Li, Dayiheng Liu, Fei Huang, Haoran Wei, et al. Qwen2.5 technical report. arXiv preprint arXiv:2412.15115, 2024.
\par
\end{samepage}

\item \begin{samepage}
Ziping Ye, Gourab Dey, Christos Christodoulopoulos, Charith Peris, Anil Ramakrishna, Weitong Ruan, Aram Galstyan, Kai-Wei Chang, Rahul Gupta, and Ninareh Mehrabi. SWAN: Semantic watermarking with abstract meaning representation. In Proceedings of the 64th Annual Meeting of the Association for Computational Linguistics (Volume 1: Long Papers), pp. 36304--36315, 2026.
\par
\end{samepage}

\item \begin{samepage}
Yifan Zhang, Zijian Wei, Haojie Ren, and Changliang Zou. e-GAI: E-value-based generalized $\alpha$-investing for online false discovery rate control. In Proceedings of the 42nd International Conference on Machine Learning, volume 267 of Proceedings of Machine Learning Research, pp. 76308--76331, 2025.
\par
\end{samepage}

\item \begin{samepage}
Xuandong Zhao, Prabhanjan Vijendra Ananth, Lei Li, and Yu-Xiang Wang. Provable robust watermarking for AI-generated text. In The Twelfth International Conference on Learning Representations, 2024.
\par
\end{samepage}

\item \begin{samepage}
Xuandong Zhao, Chenwen Liao, Yu-Xiang Wang, and Lei Li. Efficiently identifying watermarked segments in mixed-source texts. In Proceedings of the 63rd Annual Meeting of the Association for Computational Linguistics (Volume 1: Long Papers), pp. 6304--6316, 2025.
\par
\end{samepage}

\end{list}

\clearpage
\appendix
\section{Related Work}
\para{Watermarking and detection algorithms.} Recent work has substantially expanded the design space of LLM watermarking beyond early red--green token biasing (Kirchenbauer et al., 2023; 2024; Lalai et al., 2025; Liang et al., 2026). While red--green methods embed signals by increasing the probability of tokens in a key-dependent preferred set, more recent sampling-based and unbiased schemes modify the decoding process through key-dependent reweighting or sampling rules, often with the goal of preserving the original output distribution or reducing generation distortion (Mao et al., 2025; Chen et al., 2025; Huang \& Wan, 2025). Other approaches move beyond token-wise probability perturbations: structural and sequence-level methods encode watermark signals through dependencies or patterns spanning multiple tokens (Block et al., 2025; Si et al., 2026), whereas semantic and linguistic methods exploit meaning-preserving lexical, syntactic, or higher-level semantic choices (Dabiriaghdam \& Wang, 2025; Ye et al., 2026; Park et al., 2026). These watermarking schemes produce substantially different detector statistics, making localization procedures tailored to a particular statistic difficult to transfer across watermark families.

\para{Watermark Localization.} Several recent methods move beyond whole-document detection. WinMax searches over contiguous windows for the strongest standardized watermark statistic (Kirchenbauer et al., 2024), while WaterSeeker first extracts a document-wide anomaly signal and then refines suspicious regions through local traversal (Pan et al., 2025). Zhao et al.\ propose geometric-cover detection and an adaptive online locator, reducing the computational cost of span localization while treating token-level detector scores as the underlying observations (Zhao et al., 2025). Collectively, these works substantially advance where watermark evidence should be interpreted once local statistics are available. This paper pushes further by studying how to allocate a fixed query budget when obtaining local statistics incurs non-negligible cost, while retaining valid finite-sample error control.

\para{Statistical frameworks.} Watermark detection can naturally be viewed as hypothesis testing. Recent work develops pivotal-statistic frameworks for analyzing detection efficiency and optimal tests (Li et al., 2025), while adaptive sampling and thresholding bandits study how a fixed observation budget should be distributed across competing candidates (Locatelli et al., 2016). Separately, test martingales and e-values provide a framework for sequential evidence accumulation (Shafer et al., 2011; Vovk \& Wang, 2021), and e-BH controls false discoveries under arbitrary dependence (Wang \& Ramdas, 2022). These tools address complementary aspects of inference and allocation, but do not by themselves specify how heterogeneous watermark evidence should be selectively acquired and jointly tested under a detector-query budget.

\section{Background Knowledge}
This section provides the statistical and watermarking background needed for the theoretical analysis of BELL. We first restate the problem setup and notation used throughout the main text. We then review e-values, with particular emphasis on multiplicative evidence accumulation. Next, we briefly introduce four representative families of LLM watermarking methods. Finally, we review multiple hypothesis testing, false discovery rate (FDR), and the standard e-BH procedure that forms the basis of our token-weighted extension.

\subsection{Problem Setup and Notation}
Consider a document consisting of $N$ tokens
\[x_{1:N}=(x_1,\ldots,x_N),\]
partitioned into $m$ ordered contiguous blocks
\[I_1,\ldots,I_m,\qquad [m]:=\{1,\ldots,m\}.\]
For each block $I_j$, let
\[L_j:=|I_j|\]
denote its number of tokens. Since the blocks form a partition of the document,
\[\sum_{j=1}^m L_j=N.\]
Let $\theta_j\in[0,1]$ denote the watermark intensity of block $I_j$, i.e., the fraction of tokens in $I_j$ originating from watermarked generation. Given a pre-specified threshold $\theta_0$, blocks satisfying $\theta_j\leq\theta_0$ are treated as unwatermarked, whereas blocks satisfying $\theta_j>\theta_0$ are treated as watermarked. We therefore define
\[\Hnull:=\{j\in[m]:\theta_j\leq\theta_0\},\qquad\Halt:=\{j\in[m]:\theta_j>\theta_0\}.\]
Equivalently, watermark localization can be formulated as the following collection of block-level hypothesis-testing problems:
\[H_{j,0}:\theta_j\leq\theta_0,\qquad H_{j,1}:\theta_j>\theta_0,\qquad j\in[m].\]
Rejecting $H_{j,0}$ corresponds to declaring block $I_j$ as watermarked. This hypothesis-testing formulation follows the standard statistical view of detection problems (Lehmann \& Romano, 2005; Li et al., 2025).

We consider a budgeted verification setting in which the verifier can evaluate the local watermark detector at at most
\[B\leq N\]
token positions. At query round $t\in[B]$, the verifier selects a block $I_{j_t}$ and an unqueried token
\[x_{a_t}\in I_{j_t}.\]
The selection may depend on all detector evidence obtained in previous query rounds. Querying $x_{a_t}$ reveals a watermark-specific local detector signal
\[Z_{x_{a_t}}.\]
Different watermark families can produce detector signals with very different forms. To obtain a common statistical representation, BELL uses a watermark-specific adapter $\Aw$ to map the detector signal into a local e-value:
\[e_t:=e_{x_{a_t}}=\Aw\bigl(Z_{x_{a_t}}\bigr).\]
Larger values of $e_t$ represent stronger evidence against the corresponding unwatermarked null hypothesis.

For each block $j$, let
\[\mathcal T_j(B):=\{t\in[B]:j_t=j\}\]
denote the set of query rounds allocated to block $I_j$. The accumulated evidence for block $j$ after exhausting the query budget is
\[E_j(B)=\prod_{t\in\mathcal T_j(B)}e_t,\qquad E_j(0)=1,\]
where the empty product is defined as one. The terminal block-level evidence $E_1(B),\ldots,E_m(B)$ is then used to make localization decisions through multiple hypothesis testing.

\subsection{E-values and Multiplicative Evidence}
\para{E-values.} E-values provide a general way to quantify statistical evidence against a null hypothesis (Shafer et al., 2011; Vovk \& Wang, 2021; Ramdas et al., 2023; Gr\"unwald et al., 2024). For a null hypothesis $H$, a nonnegative random variable $E$ is called an e-value if
\[E\geq0,\qquad\E[E\mid H]\leq1.\]
The expectation bound, rather than a tail-probability condition, is the defining property of an e-value.

The interpretation of an e-value differs from that of a p-value. A small p-value provides evidence against the null, whereas a large e-value provides evidence against the null. In particular, values substantially larger than one indicate outcomes that would be difficult to sustain on average if the null hypothesis were true (Vovk \& Wang, 2021). An individual e-value need not be larger than one; the requirement is only that its expectation under the null is at most one.

A useful interpretation is to view an e-value as the payoff of a statistical bet. Starting from one unit of wealth, a valid e-value cannot increase the expected wealth under the null beyond one. Large realized values can therefore be interpreted as evidence obtained by a bet that would not systematically make money if the null were true (Shafer et al., 2011; Ramdas et al., 2023). This interpretation is particularly useful when evidence is accumulated over multiple observations.

A canonical example of an e-value is a likelihood ratio. Suppose a statistic $Z$ has distribution $P_0$ under the null and a reference distribution $P_1$ under an alternative, with $P_1$ absolutely continuous with respect to $P_0$. Then
\[E(Z)=\frac{dP_1}{dP_0}(Z)\]
satisfies
\[\E_{P_0}[E(Z)]=\int\frac{dP_1}{dP_0}(z)\,dP_0(z)=\int dP_1(z)=1.\]
Thus likelihood ratios, which are classical objects in hypothesis testing (Neyman \& Pearson, 1933; Lehmann \& Romano, 2005), are naturally valid e-values. More generally, likelihood-ratio constructions, exponential-moment constructions, and mixtures provide flexible ways of constructing e-values (Vovk \& Wang, 2021; Gr\"unwald et al., 2024).

\para{Multiplicative accumulation of e-values.} A key advantage of e-values is that evidence can be accumulated multiplicatively. However, it is important to distinguish between marginal validity and sequential validity. If $e_1$ and $e_2$ are merely known to satisfy
\[\E[e_1\mid H]\leq1,\qquad\E[e_2\mid H]\leq1,\]
their product $e_1e_2$ is not necessarily an e-value when the two variables are dependent. For example, the expectation of the product need not factorize into the product of their expectations.

A sufficient condition for multiplicative accumulation is that every new factor remains valid after conditioning on all information revealed before that factor is observed. Informally, for a sequence of nonnegative factors $e_1,e_2,\ldots$, we require that under the null,
\[\E[e_t\mid\text{all information available before observing }e_t]\leq1\]
at every round $t$. This condition is commonly referred to as conditional or sequential e-validity and underlies test-martingale and safe-testing constructions (Shafer et al., 2011; Vovk \& Wang, 2021; Ramdas et al., 2023; Gr\"unwald et al., 2024).

Under this condition, the product remains an e-value. To see this, consider
\[E^{(K)}:=\prod_{t=1}^K e_t.\]
Let
\[\mathcal F_{K-1}:=\text{all information available before observing }e_K.\]
Using iterated expectation,
\begin{align*}
\E\bigl[E^{(K)}\mid H\bigr]
&=\E\left[\left(\prod_{t=1}^{K-1}e_t\right)e_K\,\middle|\,H\right]\\
&=\E\left[\left(\prod_{t=1}^{K-1}e_t\right)\E[e_K\mid\mathcal F_{K-1}]\,\middle|\,H\right]\\
&\leq\E\left[\prod_{t=1}^{K-1}e_t\,\middle|\,H\right].
\end{align*}
Repeating the same argument recursively gives
\[\E\left[\prod_{t=1}^K e_t\,\middle|\,H\right]\leq1.\]
Hence
\[\prod_{t=1}^K e_t\]
is itself a valid e-value.

This result is especially useful for adaptive data acquisition. The next observation does not need to be selected independently of the previous observations. Instead, the choice of what to inspect next may depend arbitrarily on previously observed evidence; what matters is that the new factor is still conditionally valid under the null when it is obtained. This is one of the main reasons e-values are well suited to sequential and adaptive inference (Shafer et al., 2011; Howard et al., 2021; Ramdas et al., 2023).

In BELL, this principle is applied separately to each block. For a null block $j\in\Hnull$, whenever that block is queried, the newly obtained local factor is required to satisfy the corresponding conditional e-validity condition under $H_{j,0}$. Therefore, even though the rounds in $\mathcal T_j(B)$ are selected adaptively, their product
\[E_j(B)=\prod_{t\in\mathcal T_j(B)}e_t\]
remains a valid block-level e-value:
\[\E[E_j(B)\mid H_{j,0}]\leq1.\]
The formal proof of this property is provided in Appendix C.2. Conceptually, the role of the local adapter is to ensure valid evidence at each queried token, while multiplication accumulates this evidence into a single block-level statistic.

\subsection{Representative LLM Watermark Families}
LLM watermarking embeds a hidden statistical signal into model-generated text so that the signal can later be verified using the corresponding secret key or detector. Existing methods differ substantially in how this signal is introduced during generation, ranging from direct modifications of token probabilities to distribution-preserving randomized sampling (Lalai et al., 2025; Liang et al., 2026). BELL does not require all watermark families to share the same native detector statistic. Instead, it requires only that the local evidence produced by a watermark-specific detector can be mapped to a valid e-value through the adapter $\Aw$.

We briefly review the four representative watermark families considered in this work.

\para{KGW-style watermarking.} KGW-style methods embed a watermark by modifying the model's token probabilities according to a secret-key-dependent partition of the vocabulary (Kirchenbauer et al., 2023; 2024). At each generation step, the vocabulary is pseudorandomly divided into a preferred set of tokens, commonly referred to as the green list, and its complement. The logits of preferred tokens are increased before sampling, so watermarked generations contain preferred tokens more frequently than would be expected from unwatermarked text.

The resulting watermark signal is therefore based on statistical alignment between the generated sequence and a sequence of secret-key-dependent preferred token sets. This family represents watermarking schemes that deliberately introduce a small, structured bias into the next-token distribution to create a detectable signal.

\para{Unbiased reweighting watermarking.} Unbiased watermarking takes a different approach. Rather than introducing a fixed marginal bias toward a preferred subset of tokens, it uses secret-key-dependent random reweighting of the model's next-token distribution (Hu et al., 2024). Each realization of the watermark randomness changes the sampling distribution, but the original language-model distribution is recovered when averaging over the watermark randomness.

The watermark is therefore encoded in the dependence between the generated token and the secret random reweighting used during generation, rather than in a persistent shift of the marginal token distribution. This construction is attractive because it can embed a detectable keyed signal while preserving the original generation distribution in expectation.

\para{SynthID-style watermarking.} SynthID-style watermarking introduces secret-key-dependent pseudorandom scores into the sampling process (Dathathri et al., 2024). Instead of relying on a single preferred vocabulary subset, multiple keyed scoring functions can be used to rank or compare candidate tokens during generation. The sampling mechanism then favors candidates receiving favorable watermark scores.

As a result, watermarked text exhibits systematic dependence on the keyed scoring functions, whereas unwatermarked text does not exhibit the same preference. SynthID-style methods thus represent a broader class of watermarking mechanisms in which the watermark is embedded through randomized scoring and selection rules.

\para{Gumbel/exponential watermarking.} Another family constructs watermarks through secret-key-dependent randomness in the sampling procedure itself. Distortion-free watermarking methods based on exponential or equivalent Gumbel sampling exploit standard random-variable representations of categorical sampling and couple the sampling randomness to a secret key (Kuditipudi et al., 2024).

This construction can preserve the language model's marginal next-token distribution while inducing a detectable dependence between the selected token and the keyed randomness. Unlike explicit logit-bias methods, the watermark signal therefore arises from a secret coupling of the sampling process rather than from a systematic change in marginal token probabilities.

\para{Common interface used by BELL.} These four families differ substantially in both generation mechanism and native verification signal. KGW relies on keyed token preference, unbiased watermarking on random reweighting, SynthID-style methods on keyed scoring, and Gumbel/exponential methods on keyed sampling randomness. Consequently, it is generally inappropriate to assume that their raw detector outputs share a common distribution or numerical scale.

BELL separates this watermark-specific layer from the subsequent localization procedure. For a queried token $x_{a_t}$, the native detector returns $Z_{x_{a_t}}$, and the corresponding adapter converts it into
\[e_t=\Aw\bigl(Z_{x_{a_t}}\bigr).\]
Once the local evidence is represented as a valid e-value, the subsequent multiplicative accumulation and multiple-testing procedure are independent of the specific watermark family. This common e-value interface is what allows BELL to apply the same localization framework across heterogeneous watermarking mechanisms.

\subsection{Multiple Hypothesis Testing and Standard e-BH}
\para{Multiple hypothesis testing.} Classical hypothesis testing considers a single null hypothesis and decides whether the available evidence is sufficient to reject it. In multiple hypothesis testing, one instead considers a collection of hypotheses simultaneously (Lehmann \& Romano, 2005). In our localization problem, each block corresponds to one null hypothesis,
\[H_{1,0},H_{2,0},\ldots,H_{m,0},\]
and the verifier must decide which of these hypotheses to reject.

Let
\[\Shat\subseteq[m]\]
denote the set of rejected hypotheses. A block $j\in\Shat$ is reported as watermarked. Because the hypotheses are tested simultaneously, some of the rejected hypotheses may in fact be true nulls. The number of false discoveries is therefore
\[|\Shat\cap\Hnull|.\]
A central error criterion in multiple testing is the false discovery proportion (FDP), defined as
\[\FDP=\frac{|\Shat\cap\Hnull|}{|\Shat|\vee1}.\]
The denominator uses $|\Shat|\vee1$ so that FDP is defined even when no hypothesis is rejected. The false discovery rate (FDR) is the expectation of FDP:
\[\FDR=\E[\FDP]=\E\left[\frac{|\Shat\cap\Hnull|}{|\Shat|\vee1}\right].\]
Thus FDR controls the expected proportion of false discoveries among all reported discoveries.

FDR was introduced as a less conservative alternative to criteria that seek to prevent any false rejection, and has become a standard error criterion for large-scale multiple testing (Benjamini \& Hochberg, 1995). In watermark localization, FDR has a natural interpretation: among the blocks reported as watermarked, it controls the expected fraction that are actually unwatermarked.

\para{Classical BH and e-value-based multiple testing.} The classical Benjamini--Hochberg (BH) procedure performs FDR-controlling multiple testing using p-values (Benjamini \& Hochberg, 1995). Its validity depends on assumptions about the joint dependence structure of the p-values; arbitrary dependence generally requires additional correction (Benjamini \& Yekutieli, 2001).

E-values provide an alternative route to multiple testing. Because e-values are normalized through expectations rather than null-tail probabilities, they can be combined with the e-BH procedure of Wang \& Ramdas (2022). A major advantage is that e-BH controls FDR under arbitrary dependence among the input e-values.

Suppose that the $m$ hypotheses are associated with block-level e-values
\[E_1(B),E_2(B),\ldots,E_m(B),\]
such that every true null satisfies
\[\E[E_j(B)\mid H_{j,0}]\leq1,\qquad j\in\Hnull.\]
Order the e-values from largest to smallest:
\[E_{(1)}(B)\geq E_{(2)}(B)\geq\cdots\geq E_{(m)}(B),\]
where $H_{(i),0}$ denotes the null hypothesis associated with $E_{(i)}(B)$. Because larger e-values provide stronger evidence against the null, this ordering places the strongest candidates for rejection first.

For a pre-specified FDR level $\alpha\in(0,1)$, standard e-BH defines
\[i^*_{\mathrm{eBH}}=\max\left\{i\in[m]:E_{(i)}(B)\geq\frac{m}{\alpha i}\right\},\]
with
\[i^*_{\mathrm{eBH}}=0\]
if the set is empty. The procedure then rejects
\[H_{(1),0},\ldots,H_{(i^*_{\mathrm{eBH}}),0}.\]
The threshold
\[\frac{m}{\alpha i}\]
has a simple interpretation. For a candidate rejection size $i$, e-BH asks whether the $i$-th largest e-value is sufficiently large to support rejecting the top $i$ hypotheses. As $i$ increases, the threshold decreases, allowing a larger rejection set only when the ordered e-values collectively provide sufficient evidence. By taking the largest $i$ satisfying the inequality, e-BH chooses the largest self-consistent rejection set.

Indeed, if $i^*_{\mathrm{eBH}}>0$, then every rejected hypothesis satisfies
\[E_{(i)}(B)\geq E_{(i^*_{\mathrm{eBH}})}(B)\geq\frac{m}{\alpha i^*_{\mathrm{eBH}}},\qquad i\leq i^*_{\mathrm{eBH}}.\]
Thus every rejected hypothesis has an e-value exceeding the threshold determined by the final number of discoveries.

Most importantly, if the null e-values are valid, standard e-BH guarantees
\[\FDR\leq\alpha\]
without requiring independence among $E_1(B),\ldots,E_m(B)$ (Wang \& Ramdas, 2022). This robustness is particularly useful for watermark localization, because blocks belong to the same document and the query allocation strategy may depend on evidence observed from other blocks.

Standard e-BH nevertheless treats every rejected hypothesis as one discovery, regardless of the number of document tokens represented by that hypothesis. In our setting, block lengths $L_1,\ldots,L_m$ may be heterogeneous, so falsely rejecting a long block marks more unwatermarked tokens as watermarked than falsely rejecting a short block. Weighted multiple-testing methods provide a general mechanism for incorporating unequal hypothesis importance (Benjamini \& Hochberg, 1997; Genovese et al., 2006). Motivated by this mismatch between block-level FDR and token-level localization error, the main text introduces the token-weighted FDR criterion and the corresponding token-weighted e-BH procedure.

\section{Technical Details and Proof}
This section provides the technical details underlying the statistical guarantee of BELL. We first introduce two general constructions for converting a watermark-specific local detector signal into a valid e-value: the likelihood-ratio (LR) construction and the moment-generating-function (MGF) construction. We then prove the main finite-sample guarantee of BELL. The proof consists of two steps: local e-values remain valid after adaptive multiplicative accumulation within each block, and the resulting block-level e-values can be combined by token-weighted e-BH to control the token-weighted false discovery rate.

\subsection{Likelihood-ratio and MGF-based E-value Constructions}
Recall that querying token $x_{a_t}$ at round $t$ reveals a watermark-specific detector signal
\[Z_{x_{a_t}},\]
which is converted by the watermark-specific adapter $\Aw$ into a local e-value
\[e_t=\Aw\bigl(Z_{x_{a_t}}\bigr).\]
The particular form of $Z_{x_{a_t}}$ depends on the watermark family, but the subsequent localization procedure only requires the resulting $e_t$ to be valid under the corresponding null hypothesis.

Two general constructions are particularly useful for this purpose: likelihood-ratio e-values and MGF-based e-values (Neyman \& Pearson, 1933; Lehmann \& Romano, 2005; Vovk \& Wang, 2021; Wasserman et al., 2020; Gr\"unwald et al., 2024).

\para{Likelihood-ratio construction.} Consider a queried local statistic $Z$. Let $P_0$ denote its distribution under the null hypothesis and let $P_1$ be a reference alternative distribution satisfying
\[P_1\ll P_0.\]
The likelihood-ratio e-value is defined as
\[e_{\mathrm{LR}}(Z)=\frac{dP_1}{dP_0}(Z).\]
When $Z$ is discrete, this reduces to
\[e_{\mathrm{LR}}(Z)=\frac{P_1(Z)}{P_0(Z)}.\]
The distribution $P_1$ does not need to coincide with the true distribution under watermarked generation. It only specifies a reference alternative against which the observed signal is scored. The validity of the e-value depends on the normalization under $P_0$, rather than on correct specification of $P_1$.

\begin{proposition}[Validity of the LR construction]
Suppose $P_1\ll P_0$. Then
\[e_{\mathrm{LR}}(Z)=\frac{dP_1}{dP_0}(Z)\]
is an e-value under $P_0$, and in fact
\[\E_{P_0}\bigl[e_{\mathrm{LR}}(Z)\bigr]=1.\]
\end{proposition}
\begin{proof}
By the definition of the Radon--Nikodym derivative,
\begin{align*}
\E_{P_0}\bigl[e_{\mathrm{LR}}(Z)\bigr]
&=\int\frac{dP_1}{dP_0}(z)\,dP_0(z)\\
&=\int dP_1(z)\\
&=1.
\end{align*}
Therefore,
\[\E_{P_0}\bigl[e_{\mathrm{LR}}(Z)\bigr]\leq1,\]
and $e_{\mathrm{LR}}(Z)$ is a valid e-value.
\end{proof}
This is the same normalization principle underlying classical likelihood-ratio testing (Neyman \& Pearson, 1933; Lehmann \& Romano, 2005), but here the likelihood ratio is used directly as a nonnegative measure of evidence.

\para{MGF-based construction.} The LR construction requires a reference alternative distribution $P_1$. In many watermarking settings, however, the null behavior of a local detector statistic is easier to characterize than a complete alternative distribution. In this case, an e-value can instead be constructed through an exponential moment.

Let
\[T=T(Z)\]
be a real-valued statistic derived from the detector signal $Z$. For $\lambda\in\mathbb R$, define its null moment-generating function as
\[M_0(\lambda)=\E_0[\exp(\lambda T)],\]
whenever the expectation exists. Equivalently, define the null log-MGF
\[\psi_0(\lambda)=\log\E_0[\exp(\lambda T)].\]
The MGF-normalized e-value is
\[e_{\mathrm{MGF}}(Z)=\frac{\exp(\lambda T)}{M_0(\lambda)}=\exp\{\lambda T-\psi_0(\lambda)\}.\]
More generally, it is sufficient to have an upper bound
\[\psi_0^+(\lambda)\geq\log\E_0[\exp(\lambda T)].\]
In this case, we define
\[e_{\mathrm{MGF}}(Z)=\exp\{\lambda T-\psi_0^+(\lambda)\}.\]
This construction is closely related to exponential supermartingales and Chernoff-type arguments and is widely used in sequential inference (Hoeffding, 1963; Howard et al., 2021; Ramdas et al., 2023; Gr\"unwald et al., 2024).

\begin{proposition}[Validity of the MGF construction]
Suppose that
\[\psi_0^+(\lambda)\geq\log\E_0[\exp(\lambda T)].\]
Then
\[e_{\mathrm{MGF}}(Z)=\exp\{\lambda T-\psi_0^+(\lambda)\}\]
is a valid e-value under the null hypothesis.
\end{proposition}
\begin{proof}
Taking expectation under the null gives
\begin{align*}
\E_0\bigl[e_{\mathrm{MGF}}(Z)\bigr]
&=\E_0\bigl[\exp\{\lambda T-\psi_0^+(\lambda)\}\bigr]\\
&=\exp\{-\psi_0^+(\lambda)\}\E_0[\exp(\lambda T)].
\end{align*}
Since
\[\psi_0^+(\lambda)\geq\log\E_0[\exp(\lambda T)],\]
we have
\[\E_0[\exp(\lambda T)]\leq\exp\{\psi_0^+(\lambda)\}.\]
Therefore,
\[\E_0\bigl[e_{\mathrm{MGF}}(Z)\bigr]\leq1.\]
Hence $e_{\mathrm{MGF}}(Z)$ is a valid e-value.
\end{proof}
When the exact MGF is available, setting
\[\psi_0^+(\lambda)=\psi_0(\lambda)\]
gives
\[\E_0\bigl[e_{\mathrm{MGF}}(Z)\bigr]=1.\]
When the exact MGF is unavailable, any valid upper bound can be used instead, at the cost of potentially more conservative evidence.

\para{Bounded-score construction.} A useful special case follows directly from Hoeffding's lemma (Hoeffding, 1963). Suppose that
\[T\in[a,b]\]
under the null and let
\[\mu_0=\E_0[T].\]
Then
\[\log\E_0[\exp\{\lambda(T-\mu_0)\}]\leq\frac{\lambda^2(b-a)^2}{8}.\]
Consequently,
\[e(Z)=\exp\left\{\lambda(T-\mu_0)-\frac{\lambda^2(b-a)^2}{8}\right\}\]
is a valid e-value. This provides a convenient construction when only the range and null mean of a local detector statistic are known.

\para{Adaptive choice of the e-value parameter.} The tuning parameter in an LR or MGF construction may depend on detector evidence observed in earlier query rounds. For example, the reference alternative used in the LR construction or the tilting parameter $\lambda$ used in the MGF construction may be updated as more evidence is collected.

The important requirement is that the parameter used for the current query must be chosen before observing the current detector signal $Z_{x_{a_t}}$. Under this condition, the normalization arguments above continue to apply conditional on everything observed previously (Vovk \& Wang, 2021; Ramdas et al., 2023; Gr\"unwald et al., 2024). In contrast, choosing the parameter that maximizes the e-value after observing the same detector signal generally destroys the null expectation guarantee.

\para{Composite null hypotheses.} The block-level null hypothesis in our localization problem is
\[H_{j,0}:\theta_j\leq\theta_0,\]
which is generally a composite null rather than a single probability distribution. The constructions above should therefore be understood as requiring validity uniformly over the null family.

Specifically, whenever a token from a null block is queried, the local adapter must satisfy
\[\sup_{\theta\leq\theta_0}\E_\theta[e_t\mid\text{all evidence observed before the current query}]\leq1.\]
For an MGF-based construction, it is sufficient to use an upper envelope satisfying
\[\psi_0^+(\lambda)\geq\sup_{\theta\leq\theta_0}\log\E_\theta[\exp(\lambda T)].\]
Then
\[e_t=\exp\{\lambda T-\psi_0^+(\lambda)\}\]
is valid uniformly over $H_{j,0}$.

Similarly, an LR-based adapter for a composite null must be calibrated so that its expectation is at most one for every distribution included in $H_{j,0}$. A likelihood ratio constructed for only one simple null distribution does not automatically provide this uniform guarantee.

The theoretical analysis below does not otherwise depend on the particular watermark family or on whether the local adapter is constructed through an LR, an MGF, or another valid e-value construction. It uses only the local validity condition
\[\E[e_t\mid H_{j_t,0},\text{ all information available before observing }e_t]\leq1.\]

\subsection{Proof of the Main FDR Guarantee}
We now prove the main finite-sample guarantee of BELL. Recall that for each block $j$, the local e-values obtained from queries allocated to that block are accumulated as
\[E_j(B)=\prod_{t\in\mathcal T_j(B)}e_t,\qquad\mathcal T_j(B)=\{t\in[B]:j_t=j\}.\]
The proof has two parts. We first establish that $E_j(B)$ remains a valid e-value for every null block even when query allocation is adaptive. We then show that applying token-weighted e-BH to these block-level e-values controls $\FDR_L$ at the prescribed level $\alpha$.

\begin{lemma}[Validity of adaptively accumulated block evidence]
Suppose that for every null block $j\in\Hnull$, whenever round $t$ queries a token from block $j$, the newly obtained local factor satisfies
\[\E[e_t\mid H_{j,0},\text{ all evidence observed before round }t,j_t,a_t]\leq1.\]
\end{lemma}
\begin{proof}
For compactness, let $\mathcal I_t$ denote all information available immediately before observing the current local e-value $e_t$, including the previously observed evidence and the selected query $(j_t,a_t)$. For every null block $j\in\Hnull$, local e-validity requires that whenever $j_t=j$,
\[\E[e_t\mid\mathcal I_t,H_{j,0}]\leq1.\]
Define the round-$t$ multiplicative increment for block $j$ as
\[R_{j,t}=1+\1\{j_t=j\}(e_t-1).\]
Equivalently, $R_{j,t}=e_t$ if block $j$ is queried at round $t$, and $R_{j,t}=1$ otherwise. Therefore,
\[E_j(t)=E_j(t-1)R_{j,t}.\]
For any null block $j\in\Hnull$, conditioning on $\mathcal I_t$ gives
\begin{align*}
\E[R_{j,t}\mid\mathcal I_t,H_{j,0}]
&=1+\1\{j_t=j\}\bigl(\E[e_t\mid\mathcal I_t,H_{j,0}]-1\bigr)\\
&\leq1.
\end{align*}
Since $E_j(t-1)$ is determined by information available before observing $e_t$, we have
\begin{align*}
\E[E_j(t)\mid H_{j,0}]
&=\E[E_j(t-1)R_{j,t}\mid H_{j,0}]\\
&=\E\bigl[E_j(t-1)\E[R_{j,t}\mid\mathcal I_t,H_{j,0}]\mid H_{j,0}\bigr]\\
&\leq\E[E_j(t-1)\mid H_{j,0}].
\end{align*}
Applying this inequality recursively from $t=B$ to $t=1$ yields
\[\E[E_j(B)\mid H_{j,0}]\leq\E[E_j(0)\mid H_{j,0}]=1.\]
Hence $E_j(B)$ is a valid block-level e-value for every $j\in\Hnull$.

Then, for every $j\in\Hnull$,
\[\E[E_j(B)\mid H_{j,0}]\leq1.\]
Therefore, $E_j(B)$ is a valid block-level e-value.
\end{proof}

\begin{appendixtheorem}[Token-weighted FDR control]
Let
\[E_1(B),\ldots,E_m(B)\]
be the terminal block-level e-values, and suppose that
\[\E[E_j(B)\mid H_{j,0}]\leq1,\qquad j\in\Hnull.\]
No independence assumption is imposed on the joint distribution of these e-values.

Let
\[E_{(1)}(B)\geq E_{(2)}(B)\geq\cdots\geq E_{(m)}(B)\]
be the block-level e-values ordered from largest to smallest, and let $L_{(i)}$ denote the length of the block corresponding to $E_{(i)}(B)$. Define
\[C_i=\sum_{\ell=1}^i L_{(\ell)}.\]
For a prescribed level $\alpha\in(0,1)$, token-weighted e-BH selects
\[i^*_{\mathrm{TEBH}}=\max\left\{i\in[m]:E_{(i)}(B)\geq\frac{N}{\alpha C_i}\right\},\]
with $i^*_{\mathrm{TEBH}}=0$ if the set is empty.

Let $\Shat$ denote the corresponding set of rejected blocks. Then
\[\FDR_L=\E\left[\frac{\sum_{j\in\Shat\cap\Hnull}L_j}{\left(\sum_{j\in\Shat}L_j\right)\vee1}\right]\leq\alpha.\]
\end{appendixtheorem}
\begin{proof}
Define the total number of tokens contained in the rejected blocks as
\[R_L=\sum_{j\in\Shat}L_j.\]
If $R_L=0$, no block is rejected and therefore
\[\FDP_L=0.\]
The desired result is immediate in this case.

Now suppose that $R_L>0$. Let
\[k=i^*_{\mathrm{TEBH}}.\]
Because the rejected set consists of the blocks corresponding to the $k$ largest e-values,
\[R_L=C_k.\]
By the definition of token-weighted e-BH,
\[E_{(k)}(B)\geq\frac{N}{\alpha C_k}=\frac{N}{\alpha R_L}.\]
Moreover, every rejected block has an e-value at least as large as $E_{(k)}(B)$. Therefore, for every $j\in\Shat$,
\[E_j(B)\geq\frac{N}{\alpha R_L}.\]
In particular, for every falsely rejected block $j\in\Shat\cap\Hnull$,
\[1\leq\frac{\alpha R_LE_j(B)}{N}.\]
Multiplying both sides by $L_j/R_L$ gives
\[\frac{L_j}{R_L}\leq\frac{\alpha L_jE_j(B)}{N}.\]
Summing this inequality over all falsely rejected blocks yields
\begin{align*}
\FDP_L&=\frac{\sum_{j\in\Shat\cap\Hnull}L_j}{R_L}\\
&=\sum_{j\in\Shat\cap\Hnull}\frac{L_j}{R_L}\\
&\leq\frac{\alpha}{N}\sum_{j\in\Shat\cap\Hnull}L_jE_j(B)\\
&\leq\frac{\alpha}{N}\sum_{j\in\Hnull}L_jE_j(B).
\end{align*}
Taking expectations on both sides gives
\begin{align*}
\FDR_L&=\E[\FDP_L]\\
&\leq\frac{\alpha}{N}\sum_{j\in\Hnull}L_j\E[E_j(B)\mid H_{j,0}]\\
&\leq\frac{\alpha}{N}\sum_{j\in\Hnull}L_j\\
&\leq\frac{\alpha}{N}\sum_{j=1}^m L_j.
\end{align*}
Since
\[\sum_{j=1}^m L_j=N,\]
we obtain $\FDR_L\leq\alpha$. This proves the claimed token-weighted FDR control.
\end{proof}
Importantly, the proof never factorizes the joint distribution of $E_1(B),\ldots,E_m(B)$. It uses only the marginal e-value condition
\[\E[E_j(B)\mid H_{j,0}]\leq1\]
for each null block. Therefore, as in standard e-BH (Wang \& Ramdas, 2022), arbitrary dependence across the terminal block-level e-values is allowed.

\para{Completion of the main guarantee.} Lemma 1 establishes that, for every $j\in\Hnull$, the exact multiplicative block evidence produced by BELL satisfies
\[\E[E_j(B)\mid H_{j,0}]\leq1,\]
even though the allocation of the $B$ queries may depend on detector evidence observed in earlier rounds. Theorem 1 then applies directly to $E_1(B),\ldots,E_m(B)$ and yields
\[\FDR_L\leq\alpha.\]
Therefore, the adaptive allocation policy and the statistical testing layer play distinct roles. Coverage, optimistic exploration, neighborhood expansion, and near-confirmation determine where the limited query budget is spent and therefore affect localization power and query efficiency. The FDR guarantee, in contrast, depends only on two properties: the validity of the exact multiplicative block-level e-values and the terminal token-weighted e-BH rule. In particular, the clipped log-evidence and other surrogate quantities used for query allocation are not used in the final statistical test.

\section{Setup of the Experiment}
\subsection{Mixed-source Document Construction}
\para{Source documents.} We use C4 and WikiText-103 as the two source domains. All token counts are measured using the tokenizer of the corresponding generation model. For C4, we randomly sample held-out records and concatenate consecutive records with paragraph separators until the sequence is sufficiently long, after which we extract a contiguous $N=4096$-token segment. For WikiText-103, we preferentially sample contiguous text from the same Wikipedia article and concatenate neighboring articles only when a single article is too short. The resulting source sequence defines the non-generated portion of each evaluation document.

\para{Generated-region lengths and locations.} Each positive document contains five non-overlapping model-generated regions, denoted by
\[\mathcal R_1,\ldots,\mathcal R_5.\]
Importantly, the range 800--1200 refers to the combined length of the five generated regions, rather than the length of each individual region. Specifically, we independently sample
\[R_r\sim\operatorname{Uniform}\bigl(\{160,164,\ldots,236,240\}\bigr),\qquad r=1,\ldots,5,\]
where $R_r=|\mathcal R_r|$. Therefore,
\[R_{\mathrm{gen}}=\sum_{r=1}^5R_r\in[800,1200].\]
Restricting $R_r$ to multiples of four allows the partial-provenance patterns described below to realize the prescribed watermark intensities exactly.

To ensure that the watermark signal is distributed across the document rather than concentrated in a single neighborhood, we divide the 4096-token document into five coarse ordered placement intervals,
\[\mathcal D_r=\left[\left\lfloor\frac{(r-1)N}{5}\right\rfloor+1,\left\lfloor\frac{rN}{5}\right\rfloor\right],\qquad r=1,\ldots,5.\]
One generated region $\mathcal R_r$ is placed inside each $\mathcal D_r$. The starting positions are sampled uniformly subject to three constraints: (i) each generated region lies completely inside its corresponding placement interval; (ii) the five generated regions do not overlap; and (iii) two consecutive generated regions are separated by at least
\[g_{\min}=128\]
original source tokens. If a sampled placement violates these constraints, the locations are resampled.

This construction spreads the five generated regions across different portions of the document while retaining explicit ungenerated gaps between them. Region locations are sampled independently of the block partition introduced below.

\para{Replacement-based construction.} Rather than inserting additional text into the source document, we replace the source passage corresponding to each $\mathcal R_r$ with a model-generated passage of exactly the same length $R_r$. Consequently, the final document always contains exactly $N=4096$ tokens.

Generation within each region is autoregressive. The same model, decoding configuration, and generation procedure are used throughout the region. For partially watermarked regions, the model state and autoregressive context remain continuous when the watermark mechanism is switched on or off; only the watermark transformation changes according to the provenance mask defined below.

\para{Region-level watermark intensities.} The five generated regions have the fixed multiset of watermark intensities
\[\Theta_{\mathrm{reg}}=\{1.0,1.0,0.75,0.75,0.50\}.\]
Thus, every positive document contains two fully watermarked regions, two regions with watermark intensity 0.75, and one region with watermark intensity 0.50.

For each document, we randomly permute these five intensities across the five spatial locations $\mathcal R_1,\ldots,\mathcal R_5$. Hence watermark strength is not systematically associated with an absolute document position.

Let
\[z_i\in\{0,1\}\]
denote the token-level provenance indicator, where $z_i=1$ means that token $i$ was generated with the watermark mechanism enabled and $z_i=0$ means that the watermark mechanism was disabled. Tokens belonging to the original source background also have $z_i=0$.

\para{Partial provenance within a generated region.} To realize different watermark intensities while preserving a continuous generated passage, each generated region $\mathcal R_r$ is divided into four consecutive equal-length micro-regions,
\[\mathcal R_{r,1},\mathcal R_{r,2},\mathcal R_{r,3},\mathcal R_{r,4},\]
each containing $R_r/4$ tokens. We use the following provenance patterns:
\begin{align*}
\theta=1.00&:\qquad\mathrm W-\mathrm W-\mathrm W-\mathrm W,\\
\theta=0.75&:\qquad\mathrm W-\mathrm W-\mathrm W-\mathrm U,\\
\theta=0.50&:\qquad\mathrm W-\mathrm U-\mathrm W-\mathrm U,
\end{align*}
where $\mathrm W$ denotes generation with the watermark transformation enabled and $\mathrm U$ denotes generation with the watermark transformation disabled.

Because $R_r$ is always divisible by four, these patterns realize their nominal region-level intensities exactly:
\[\frac{1}{R_r}\sum_{i\in\mathcal R_r}z_i=\begin{cases}1,&\theta=1.00,\\0.75,&\theta=0.75,\\0.50,&\theta=0.50.\end{cases}\]
The autoregressive generation context remains continuous across all four micro-regions. In particular, the $\mathrm U$ portions are not independently generated passages; they are generated by the same model trajectory with only the watermark transformation switched off.

\para{Variable-length block partition.} After the complete 4096-token mixed-source document has been constructed, we partition it into
\[m=40\]
ordered contiguous blocks
\[I_1,\ldots,I_{40}.\]
The block partition is sampled independently of both the generated-region boundaries and the internal provenance boundaries. Consequently, generated regions generally span multiple blocks, and blocks near a boundary can contain mixtures of watermarked, unwatermarked generated, and original source tokens.

To produce heterogeneous block lengths, we draw independent positive log-normal weights
\[\widetilde L_j\sim\operatorname{LogNormal}(0,0.35^2),\qquad j=1,\ldots,40,\]
and rescale them according to
\[L_j^{(0)}=N\frac{\widetilde L_j}{\sum_{\ell=1}^{40}\widetilde L_\ell}.\]
The resulting values are rounded to integer token lengths and constrained to
\[48\leq L_j\leq256.\]
Any residual difference caused by rounding and clipping is distributed across eligible blocks so that the final partition satisfies
\[\sum_{j=1}^{40}L_j=4096.\]
This preserves substantial variation in block length while keeping the average block size close to $4096/40\approx102.4$ tokens.

\para{Block-level watermark intensity and ground truth.} The localization hypotheses are defined at the block level rather than at the generated-region level. After document construction and block partitioning, the watermark intensity of block $I_j$ is computed directly from the token-level provenance mask:
\[\theta_j=\frac{1}{L_j}\sum_{i\in I_j}z_i.\]
Because block boundaries are independent of both generated-region boundaries and internal watermark-state boundaries, $\theta_j$ may take intermediate values even though the nominal region-level intensities are restricted to 1.0, 0.75, and 0.50.

Following the experimental setup in the main text, we use
\[\theta_0=0.6.\]
A block is treated as a watermark localization target when
\[\theta_j>\theta_0,\]
and as a null block when
\[\theta_j\leq\theta_0.\]
Thus, ground-truth block labels are always determined from the realized token-level provenance mask rather than directly inherited from the nominal intensity of the generated region with which a block overlaps.

\para{Summary.} Each evaluation document contains exactly 4096 tokens and is partitioned into 40 variable-length blocks. It contains five spatially separated generated regions whose combined length is between 800 and 1200 tokens. The five region-level watermark intensities consist of exactly two 1.0 regions, two 0.75 regions, and one 0.50 region, randomly assigned to the five document locations. The generated-region boundaries, internal watermark-state boundaries, and block boundaries are not aligned. This construction produces multiple separated watermark-bearing intervals, heterogeneous watermark strengths, and mixed-provenance boundary blocks, providing a controlled setting for evaluating both localization accuracy and query efficiency.

\subsection{Hyper-parameters of BELL}
Table 3 summarizes the hyper-parameters for BELL in the experiment.
\begin{table}[H]
\centering
\caption{Allocation parameters in our BELL mechanism-level implementation.}
\label{tab:3}
\vspace{4pt}
\begin{tabular}{@{}ll@{}}
\toprule
Name & Value \\
\midrule
coverage exponent $\beta$ & 0.50 \\
minimum initial coverage $b_{\min}$ & 4 queries per block \\
coverage scale $c_{\mathrm{cov}}$ & 0.35 \\
forced-coverage rounds $\mathcal T_{\mathrm{cov}}$ & every 25 rounds after initial coverage \\
exploration parameter $\delta$ & 0.05 \\
neighborhood weight $\eta$ & 0.50 \\
planning fraction $\kappa$ & 0.25 \\
confirmation horizon $H$ & 64 queries \\
planning denominator $\varepsilon$ & 0.05 \\
allocation clipping $C_-,C_+$ & 0.8, 0.8 \\
terminal FDR level $\alpha$ & 0.05 \\
number of blocks $m$ & 40 \\
\bottomrule
\end{tabular}
\end{table}

\subsection{Adapting Baselines to the Budgeted Setting}
WinMax (Kirchenbauer et al., 2024), WaterSeeker (Pan et al., 2025), and AOL (Zhao et al., 2025) were designed for the full-scan setting. They compute the watermark score of every token and localize watermarked segments from this dense score sequence. In our setting, the detector can be evaluated at no more than $B$ token positions. We therefore keep the localization rule of each method and change only how token scores are obtained. Each baseline evaluates the detector on a sparse subset of $B$ positions instead of all $N$. It computes its statistics from the scores of the queried tokens only, and unqueried tokens are ignored rather than imputed. The budget fraction $B/N$ thus determines how densely each baseline scans the document, and $B=N$ recovers the original full-scan method. We refer to the adapted methods as B-WinMax, B-WaterSeeker, and B-AOL.

The three adaptations share a coarse-to-fine pattern. Part of the budget is spent on sparse probes spread evenly over the document. The rest is spent on positions that the method itself identifies as promising from these probes. The methods differ in how they identify those positions.

\para{B-WinMax.} WinMax scans windows of varying lengths and returns the window with the largest z-score. B-WinMax first probes the document at evenly spaced positions and computes each window's z-score from the probed tokens inside it. It then spends the remaining budget on unqueried tokens in and around the highest-scoring windows, and recomputes the window scores as new tokens are queried. When the budget is exhausted, it reports the windows selected by the WinMax criterion from the sparse scores.

\para{B-WaterSeeker.} WaterSeeker first smooths the token scores to detect suspicious regions, and then traverses each region locally to determine its boundaries. B-WaterSeeker runs the first stage on sparse probes spread over the document. It spends the remaining budget on unqueried tokens within and around the suspicious regions. The local traversal is then carried out on the queried tokens, and the resulting segments are reported.

\para{B-AOL.} AOL treats the token scores as a noisy sequence and estimates its local mean with an adaptive online estimator. It reports the positions whose estimate exceeds a threshold. B-AOL applies the same estimator to the queried tokens in document order. It places subsequent queries in regions whose current estimate is high or close to the threshold, which refines the estimated segment boundaries. When the budget is exhausted, it reports the segments given by the final estimate.

In all three cases, the watermark detector and the cost of a query are the same as for BELL. Every method therefore evaluates the detector the same number of times at a given $B/N$. The methods differ only in where they place their queries and in how they turn the queried scores into localized segments.

\subsection{Adapter Construction}
We now discuss how watermark-specific detector outputs can be converted into e-evidence that is compatible with the statistical layer of BELL. A useful distinction is between the pure-null case $\theta_0=0$ and the general composite-null case $\theta_0>0$.

When $\theta_0=0$, every token in a null block is unwatermarked. In this case, standard likelihood-ratio or exponential-moment constructions can be applied directly at the token level. When $\theta_0>0$, however, the block-level null hypothesis itself permits a nonzero fraction of watermarked tokens. Consequently, a token queried from a null block need not be an unwatermarked token, and a token-level e-value calibrated only under $z_i=0$ is no longer sufficient. In this regime, the adapter should be calibrated to the block-level composite null and may naturally depend on $\theta_0$.

We first describe the standard adapters for $\theta_0=0$, and then discuss several general strategies for constructing $\theta_0$-dependent adapters.

\para{Standard token-level adapters when $\theta_0=0$.} Let $\mathcal G_t$ denote all information available before the detector score of the current queried position is revealed, including all previously observed detector outputs and any round-$t$ tuning parameters. For watermark family $W$, let $s_i^{W}$ denote the scheme-specific identifier determining the keyed randomization used to score position $i$. This identifier may correspond, for example, to a context-dependent seed, a context code, or a key-coordinate index.

To avoid treating repeated keyed randomness as fresh statistical evidence, we associate each position with an eligibility indicator $\chi_i^{W}\in\{0,1\}$. The eligibility indicator is determined before the corresponding keyed detector score is inspected. Positions for which the required adapter inputs are unavailable, or for which the same keyed randomization has already been used under the scoring rule, are assigned the neutral factor
\begin{equation}e_{i,t}^{W}=1.\end{equation}
For eligible positions, the constructions below satisfy
\begin{equation}
\E[e_{i,t}^{W}\mid\mathcal G_t,a_t=i,z_i=0,\chi_i^{W}=1]\leq1.
\end{equation}
All tuning parameters appearing below are either fixed before testing or are $\mathcal G_t$-measurable, i.e., they are chosen before observing the current detector score.

Under $\theta_0=0$, a null block contains only positions with $z_i=0$. Therefore, the conditional validity in Eq.\ (10) is sufficient for multiplicative evidence accumulation.

\para{KGW-style watermarking.} Let
\begin{equation}Z_i^{\mathrm{KGW}}:=\1\{x_i\in G_i\},\end{equation}
where $G_i$ is the keyed green list associated with position $i$. For an eligible unwatermarked position, the null randomization satisfies
\begin{equation}
Z_i^{\mathrm{KGW}}\mid\mathcal G_t,a_t=i,z_i=0,\chi_i^{\mathrm{KGW}}=1\sim\operatorname{Bernoulli}(\gamma),
\end{equation}
where $\gamma\in(0,1)$ is the green-list inclusion probability.

For any predictable reference parameter $p_t^\star\in(\gamma,1)$, define
\begin{equation}
e_{i,t}^{\mathrm{KGW}}:=\left(\frac{p_t^\star}{\gamma}\right)^{Z_i^{\mathrm{KGW}}}\left(\frac{1-p_t^\star}{1-\gamma}\right)^{1-Z_i^{\mathrm{KGW}}}.
\end{equation}
This is a likelihood-ratio e-value, and
\begin{equation}
\E[e_{i,t}^{\mathrm{KGW}}\mid\mathcal G_t,a_t=i,z_i=0,\chi_i^{\mathrm{KGW}}=1]=1.
\end{equation}
The reference parameter $p_t^\star$ need not equal the true watermarked green-token probability; misspecification affects power but not null validity.

\para{Hu-style unbiased reweighting.} Let $P_i$ denote the reference next-token distribution and $Q_i$ the corresponding keyed reweighted distribution. For an eligible position, the unbiased-reweighting property gives
\begin{equation}
\E[Q_i(x)\mid\mathcal G_t,a_t=i,x_i,P_i,z_i=0,\chi_i^{\mathrm{Hu}}=1]=P_i(x)
\end{equation}
for every token $x$. We therefore define
\begin{equation}
e_{i,t}^{\mathrm{Hu}}:=\frac{Q_i(x_i)}{P_i(x_i)},\qquad P_i(x_i)>0.
\end{equation}
It follows immediately that
\begin{equation}
\E[e_{i,t}^{\mathrm{Hu}}\mid\mathcal G_t,a_t=i,x_i,P_i,z_i=0,\chi_i^{\mathrm{Hu}}=1]=1.
\end{equation}
Positions for which $P_i(x_i)=0$ or the required adapter inputs are unavailable are assigned the neutral factor in Eq.\ (9).

\para{SynthID-style watermarking.} Consider an eligible position with $K_{\mathrm{syn}}$ binary keyed layers
\[g_{i,1},\ldots,g_{i,K_{\mathrm{syn}}}.\]
Under the null randomization,
\begin{equation}
(g_{i,1},\ldots,g_{i,K_{\mathrm{syn}}})\mid\mathcal G_t,a_t=i,z_i=0,\chi_i^{\mathrm{Syn}}=1\sim\operatorname{Bernoulli}(1/2)^{\otimes K_{\mathrm{syn}}}.
\end{equation}
For any predictable $\lambda_t\geq0$, define
\begin{equation}
e_{i,t}^{\mathrm{Syn}}:=\exp\left\{\lambda_t\sum_{k=1}^{K_{\mathrm{syn}}}g_{i,k}-K_{\mathrm{syn}}\log\left(\frac{1+\exp(\lambda_t)}{2}\right)\right\}.
\end{equation}
The null moment-generating function gives
\begin{equation}
\E[e_{i,t}^{\mathrm{Syn}}\mid\mathcal G_t,a_t=i,z_i=0,\chi_i^{\mathrm{Syn}}=1]=1.
\end{equation}
The $K_{\mathrm{syn}}$ keyed layers are treated jointly as one token-level observation.

\para{Gumbel/exponential watermarking.} Let $U_{i,x_i}$ denote the keyed uniform coordinate used by the local exponential-minimum detector, and define
\begin{equation}Z_i^{\mathrm{Exp}}:=-\log U_{i,x_i}.\end{equation}
For an eligible unwatermarked position,
\begin{equation}
U_{i,x_i}\mid\mathcal G_t,a_t=i,z_i=0,\chi_i^{\mathrm{Gum}}=1\sim\operatorname{Unif}(0,1),
\end{equation}
and hence
\[Z_i^{\mathrm{Exp}}\sim\operatorname{Exp}(1).\]
For any predictable $\lambda_t\geq0$, define
\begin{equation}
e_{i,t}^{\mathrm{Gum}}:=(1+\lambda_t)\exp\{-\lambda_t Z_i^{\mathrm{Exp}}\}.
\end{equation}
Since
\[\E[\exp\{-\lambda_t Z_i^{\mathrm{Exp}}\}]=\frac{1}{1+\lambda_t},\]
we obtain
\begin{equation}
\E[e_{i,t}^{\mathrm{Gum}}\mid\mathcal G_t,a_t=i,z_i=0,\chi_i^{\mathrm{Gum}}=1]=1.
\end{equation}

\para{Accumulation under the pure null.} For $\theta_0=0$, every position in a null block satisfies $z_i=0$. Therefore, if block $j$ is queried at rounds
\[\mathcal T_j(t):=\{s\leq t:j_s=j\},\]
the product
\begin{equation}E_j(t):=\prod_{s\in\mathcal T_j(t)}e_s\end{equation}
is a nonnegative e-process under $H_{j,0}$, provided the local conditional-validity condition is maintained at every query. Consequently,
\begin{equation}\E[E_j(\tau)\mid H_{j,0}]\leq1\end{equation}
for any bounded stopping time $\tau$ generated by the adaptive query history.

\para{Why $\theta_0>0$ is different.} The situation changes fundamentally when the block-level null is
\begin{equation}H_{j,0}:\theta_j\leq\theta_0,\qquad\theta_0>0.\end{equation}
Writing
\begin{equation}K_j:=\lfloor\theta_0L_j\rfloor,\end{equation}
the null now permits as many as $K_j$ genuinely watermarked positions inside block $j$. Thus, even when $H_{j,0}$ is true, a queried position may satisfy $z_i=1$.

This observation is important because the standard adapters in Eqs.\ (13)--(23) are calibrated under $z_i=0$. In general,
\[\E[e_{i,t}^{W}\mid z_i=1]\]
may be substantially larger than one. Therefore, simply multiplying the $\theta_0=0$ token-level e-values does not in general yield a valid e-value for the composite null in Eq.\ (27).

More generally, there is no reason to require that every individual query be an e-value for the entire block-level composite null. For $\theta_0>0$, it is often more natural to construct a $\theta_0$-dependent block-level e-process
\begin{equation}
E_{j,t}^{(\theta_0)}=\mathcal A_{W}^{(\theta_0)}(Z_{j,1:t};L_j,K_j),
\end{equation}
whose validity requirement is directly
\begin{equation}\sup_{\theta_j\leq\theta_0}\E[E_{j,\tau}^{(\theta_0)}]\leq1.\end{equation}
Here $\tau$ may be the adaptively determined number of queries allocated to block $j$. The allocation layer of BELL can still use local detector scores or predictable evidence surrogates to decide where to query; only the terminal block-level quantity used by e-BH needs to satisfy Eq.\ (30).

Different choices of $\theta_0$ therefore naturally induce different adapters. Below we outline several useful constructions.

\para{Insight 1: finite-population likelihood adapters.} Suppose that a watermark-specific score $Y$ admits a conditional model
\begin{equation}Y\mid z=0\sim P_0,\qquad Y\mid z=1\sim P_1,\end{equation}
or, more generally, $P_1$ belongs to a specified family $\mathcal P_1$.

Assume queried locations within a selected block are sampled uniformly without replacement. Conditional on a block containing exactly $k$ watermarked positions, the latent provenance sequence of the queried tokens then follows a finite-population sampling law. Let
\begin{equation}p_{j,t}^{(k,\eta)}(Y_{1:t})\end{equation}
denote the likelihood of the first $t$ queried scores when block $j$ contains exactly $k$ watermarked positions and $\eta$ denotes any nuisance parameter governing the watermarked-score distribution.

A reference alternative can be formed by mixing over $k>K_j$:
\begin{equation}
Q_{j,t}:=\sum_{\ell=K_j+1}^{L_j}\pi_\ell p_{j,t}^{(\ell,\eta_\ell)},\qquad\sum_{\ell=K_j+1}^{L_j}\pi_\ell=1.
\end{equation}
A composite-null e-value is then obtained by
\begin{equation}
E_{j,t}^{\mathrm{LR}}:=\frac{Q_{j,t}}{\displaystyle\sup_{\substack{0\leq k\leq K_j\\\eta\in\mathcal H_0}}p_{j,t}^{(k,\eta)}}.
\end{equation}
For every null model $(k_0,\eta_0)$,
\begin{equation}E_{j,t}^{\mathrm{LR}}\leq\frac{Q_{j,t}}{p_{j,t}^{(k_0,\eta_0)}}.\end{equation}
The quantity on the right is a likelihood-ratio process under the corresponding simple null. Hence Eq.\ (34) is valid uniformly over the composite null, provided the null family contains the true score law.

This construction is attractive because it treats $\theta_j\leq\theta_0$ directly as a finite-population hypothesis rather than repeatedly protecting each individual query against the worst possible provenance assignment. The likelihood $p_{j,t}^{(k,\eta)}$ can be evaluated by a simple dynamic program over the latent number of queried watermarked positions.

\para{Insight 2: finite-population MGF and bounded-score adapters.} Exact likelihood models for watermarked positions may not always be available. A weaker alternative is to characterize the moment-generating behavior of a score statistic $T$.

Suppose that, for a fixed $\lambda$,
\begin{align}
\log\E[\exp(\lambda T)\mid z=0]&\leq\psi_0^+(\lambda),\\
\log\E[\exp(\lambda T)\mid z=1]&\leq\psi_1^+(\lambda).
\end{align}
Consider a uniformly sampled batch of $h$ positions from a block of length $L_j$. If the block contains exactly $k$ watermarked positions, the number $S$ of watermarked positions in the batch follows
\begin{equation}S\sim\operatorname{Hypergeom}(L_j,k,h).\end{equation}
Hence a valid composite-null normalizer is
\begin{equation}
\begin{split}
M_{j,h}^+(\lambda):={}&\max_{0\leq k\leq K_j}\sum_s\frac{\binom{k}{s}\binom{L_j-k}{h-s}}{\binom{L_j}{h}}\\
&\quad\times\exp\{s\psi_1^+(\lambda)+(h-s)\psi_0^+(\lambda)\}.
\end{split}
\end{equation}
This yields the batch-level e-value
\begin{equation}E_{j,h}^{\mathrm{MGF}}:=\frac{\exp\{\lambda\sum_{r=1}^hT_r\}}{M_{j,h}^+(\lambda)}.\end{equation}
A particularly robust special case requires no explicit distributional model for watermarked positions. Suppose a nonnegative raw factor $R_i$ satisfies
\begin{equation}\E[R_i\mid z_i=0]\leq1,\qquad0\leq R_i\leq c\end{equation}
for some finite $c\geq1$. For a uniformly sampled batch of size $h$, define
\begin{equation}C_{j,h}(c):=\sum_s\frac{\binom{K_j}{s}\binom{L_j-K_j}{h-s}}{\binom{L_j}{h}}c^s,\end{equation}
where the sum is over feasible $s$. Then
\begin{equation}E_{j,h}^{\mathrm{cap}}:=\frac{\prod_{r=1}^hR_r}{C_{j,h}(c)}\end{equation}
is valid under any fixed provenance mask satisfying $\sum_{i\in I_j}z_i\leq K_j$.

The finite-population normalization in Eq.\ (42) uses the fact that the null has a global budget of at most $K_j$ watermarked positions. This can be considerably less conservative than repeatedly treating every query as if the full contamination budget were still available. The batch size and any tuning parameters must be chosen before observing the scores in that batch.

\para{Insight 3: mixtures can adapt to unknown watermark strength.} The alternative watermark intensity is typically unknown. Even when $\theta_0$ is fixed, blocks under the alternative may have very different values of $\theta_j$. An adapter therefore need not commit to a single alternative intensity or a single tuning parameter.

For example, the alternative likelihood in Eq.\ (33) may place mass on several candidate intensities,
\begin{equation}\pi_\ell\propto\pi(\ell/L_j),\qquad\ell>K_j,\end{equation}
so that the resulting e-value accumulates evidence across a range of possible watermark strengths.

Similarly, if $E_{j,t}(\lambda)$ is a valid e-value for every fixed $\lambda$, then a fixed or predictable mixture
\begin{equation}E_{j,t}^{\mathrm{mix}}:=\int E_{j,t}(\lambda)\,d\Pi(\lambda)\end{equation}
remains valid. Such mixtures can reduce sensitivity to a single hand-tuned value of $p^\star$ or $\lambda$ while preserving the statistical guarantee.

More generally, the amount of structure imposed on the watermarked-position distribution determines a robustness--power tradeoff. A large null family $\mathcal P_1$ provides stronger robustness but typically yields more conservative evidence. A more informative family can substantially improve power, provided that the assumed family still contains the true score distribution.

\para{Summary.} The standard adapters in Eqs.\ (13)--(23) provide a simple and general interface when $\theta_0=0$. For $\theta_0>0$, however, the null hypothesis permits genuine watermarked tokens and therefore requires a $\theta_0$-aware calibration.

The key point is that BELL does not fundamentally require every queried token to produce an individually valid e-value. What is required by the terminal multiple-testing layer is a valid block-level e-value,
\begin{equation}\sup_{\theta_j\leq\theta_0}\E[E_j(B)]\leq1.\end{equation}
Finite-population likelihood ratios, MGF-based constructions, bounded score calibrations, and mixtures provide several principled ways to construct such adapters. The appropriate choice depends on how much is known about the detector-score distribution for the corresponding watermark family.

\clearpage
\section{Supplementary Experimental Results}
We provide the supplementary results of our experiments in this section. The proposed algorithm BELL consistently outperforms the existing baselines across different watermark families, datasets, and LLMs.

\subsection{The Query Effectiveness Comparison}
To complement the query-efficiency results reported in Figure 1, we provide the corresponding comparisons for the remaining model--dataset configurations. For each watermark family, we report the minimum query fraction $B/N$ required by different localization methods to achieve target recall levels ranging from 70\% to 90\%. Lower query fractions indicate higher query efficiency.

\figpaperhere{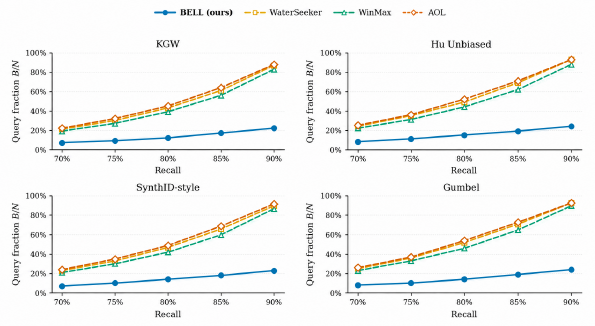}{0.80}{Comparison of the minimum query fraction $B/N$ required to achieve different target recall levels across various algorithms. All results are evaluated using Qwen2.5-7B-Instruct on the WikiText-103 dataset.}{fig:3}
\figpaperhere{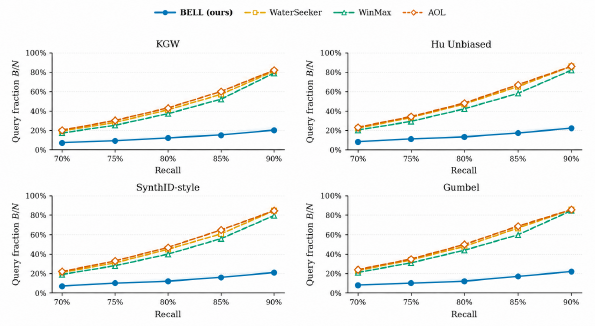}{0.80}{Comparison of the minimum query fraction $B/N$ required to achieve different target recall levels across various algorithms. All results are evaluated using Mistral-7B-Instruct-v0.3 on the C4 dataset.}{fig:4}
\figpaperhere{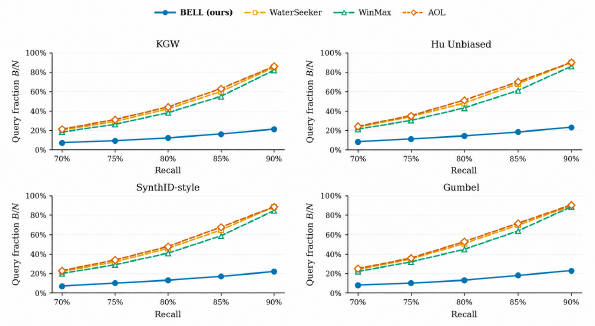}{0.80}{Comparison of the minimum query fraction $B/N$ required to achieve different target recall levels across various algorithms. All results are evaluated using Mistral-7B-Instruct-v0.3 on the WikiText-103 dataset.}{fig:5}

Across all three additional configurations, BELL consistently requires a substantially smaller query fraction than WaterSeeker, WinMax, and AOL to reach the same target recall. The advantage becomes particularly pronounced at higher recall levels: at 90\% recall, BELL requires only about 20\%--24\% of the document tokens to be queried across the considered settings, whereas the competing methods generally require approximately 79\%--93\%. These results show that the query-efficiency improvement of BELL is consistent across different LLM backbones, source datasets, and watermark families.

\subsection{The \texorpdfstring{$\FDR_L$}{FDR-L} Control Verification}
We further empirically verify the token-weighted false discovery rate ($\FDR_L$) control under a wide range of experimental conditions. Specifically, we evaluate four localization algorithms under four watermark families, four LLM--dataset configurations, and five query fractions $B/N\in\{5\%,10\%,20\%,30\%,50\%\}$, resulting in a total of $4\times4\times4\times5=320$ $\FDR_L$ measurements.

Table 4 summarizes the results by localization algorithm and watermark family. For each algorithm--family pair, the minimum, maximum, and average $\FDR_L$ are computed over all four LLM--dataset configurations and all five query fractions, i.e., 20 experimental conditions. All values are reported as percentages.
\begin{table}[H]
\centering
\caption{Empirical $\FDR_L$ statistics (\%) across different watermark families. For each algorithm--family pair, Min, Max, and Avg are computed over four LLM--dataset configurations and five query fractions (20 conditions in total).}
\label{tab:4}
\vspace{4pt}
\resizebox{\linewidth}{!}{%
\begin{tabular}{@{}l*{12}{c}@{}}
\toprule
\multirow{2}{*}{Algorithm} & \multicolumn{3}{c}{KGW} & \multicolumn{3}{c}{Hu Unbiased} & \multicolumn{3}{c}{SynthID-style} & \multicolumn{3}{c}{Gumbel} \\
\cmidrule(lr){2-4}\cmidrule(lr){5-7}\cmidrule(lr){8-10}\cmidrule(lr){11-13}
 & Min & Max & Avg & Min & Max & Avg & Min & Max & Avg & Min & Max & Avg \\
\midrule
WaterSeeker & 2.21 & 3.19 & 2.76 & 2.70 & 3.52 & 3.05 & 2.35 & 3.44 & 2.92 & 2.72 & 3.55 & 3.25 \\
WinMax & 2.11 & 2.93 & 2.62 & 2.29 & 3.39 & 2.74 & 2.08 & 3.15 & 2.74 & 2.33 & 3.28 & 2.93 \\
AOL & 2.71 & 3.80 & 3.33 & 2.88 & 3.99 & 3.45 & 2.92 & 3.75 & 3.44 & 3.05 & 4.08 & 3.63 \\
\midrule
\textbf{BELL (ours)} & \textbf{0.79} & \textbf{1.76} & \textbf{1.36} & \textbf{1.10} & \textbf{2.05} & \textbf{1.61} & \textbf{1.03} & \textbf{1.98} & \textbf{1.48} & \textbf{1.14} & \textbf{2.09} & \textbf{1.68} \\
\bottomrule
\end{tabular}
}
\end{table}

Across all 320 measurements, every empirical $\FDR_L$ remains below the prescribed significance level $\alpha=0.05$. The largest observed value over all algorithms, watermark families, LLM--dataset configurations, and query fractions is 4.08\%, which is still below the 5\% threshold. This observation also holds consistently across the full range of query budgets, indicating that the error-control behavior is stable under different levels of query expenditure.

For BELL, the empirical $\FDR_L$ values are substantially below the prescribed level across all four watermark families. In particular, its maximum $\FDR_L$ ranges from only 1.76\% to 2.09\%, while its average $\FDR_L$ ranges from 1.36\% to 1.68\%. These empirical results are consistent with Theorem 3.3, which establishes that the proposed token-weighted e-BH procedure controls $\FDR_L$ at the prescribed level. Thus, the experiments provide complementary empirical verification of the theoretical error-control guarantee across different watermark families, LLM backbones, source datasets, and query budgets.

\subsection{The Comprehensive Comparison with Baselines}
We provide the complete comparison with the baseline methods for the remaining model--dataset configurations. Following Table 1 in the main text, Precision, Recall, F1, and IoU are evaluated at a fixed query fraction of $B/N=20\%$. We additionally report $B^\star(0.8)$, which denotes the minimum query fraction required to reach a target recall of 80\% while satisfying the prescribed FDR constraint. All reported values are percentages.
\begin{table}[H]
\centering
\caption{Comparison of different methods at a fixed query fraction of $B/N=20\%$ with various watermarking schemes. All results are evaluated using Qwen2.5-7B-Instruct with WikiText-103.}
\label{tab:5}
\vspace{4pt}
\resizebox{\linewidth}{!}{%
\begin{tabular}{@{}l*{10}{c}@{}}
\toprule
\multirow{2}{*}{Method} & \multicolumn{5}{c}{KGW} & \multicolumn{5}{c}{Hu Unbiased} \\
 & Prec.\,$\uparrow$ & Rec.\,$\uparrow$ & F1\,$\uparrow$ & IoU\,$\uparrow$ & $B^{\star}(0.8)\,\downarrow$ & Prec.\,$\uparrow$ & Rec.\,$\uparrow$ & F1\,$\uparrow$ & IoU\,$\uparrow$ & $B^{\star}(0.8)\,\downarrow$ \\
\midrule
WaterSeeker & 83.7 & 67.3 & 74.6 & 59.5 & 43 & 82.2 & 59.2 & 68.8 & 52.4 & 49 \\
WinMax & 85.6 & 71.0 & 77.6 & 63.4 & 39 & 85.6 & 63.5 & 72.9 & 57.4 & 44 \\
AOL & 85.2 & 63.2 & 72.6 & 57.0 & 45 & 82.9 & 56.6 & 67.3 & 50.7 & 52 \\
\midrule
\textbf{BELL (ours)} & \textbf{95.0} & \textbf{88.6} & \textbf{91.7} & \textbf{84.7} & \textbf{12} & \textbf{94.3} & \textbf{87.1} & \textbf{90.6} & \textbf{82.7} & \textbf{15} \\
\bottomrule
\addlinespace[7pt]
\toprule
\multirow{2}{*}{Method} & \multicolumn{5}{c}{SynthID-style} & \multicolumn{5}{c}{Gumbel} \\
 & Prec.\,$\uparrow$ & Rec.\,$\uparrow$ & F1\,$\uparrow$ & IoU\,$\uparrow$ & $B^{\star}(0.8)\,\downarrow$ & Prec.\,$\uparrow$ & Rec.\,$\uparrow$ & F1\,$\uparrow$ & IoU\,$\uparrow$ & $B^{\star}(0.8)\,\downarrow$ \\
\midrule
WaterSeeker & 82.9 & 60.6 & 70.0 & 53.8 & 47 & 82.2 & 56.7 & 67.1 & 50.5 & 52 \\
WinMax & 84.7 & 67.7 & 75.3 & 60.4 & 42 & 85.5 & 60.1 & 70.6 & 54.6 & 46 \\
AOL & 84.7 & 58.3 & 69.1 & 52.8 & 49 & 84.8 & 53.5 & 65.6 & 48.8 & 54 \\
\midrule
\textbf{BELL (ours)} & \textbf{94.2} & \textbf{88.1} & \textbf{91.0} & \textbf{83.6} & \textbf{14} & \textbf{94.6} & \textbf{86.6} & \textbf{90.4} & \textbf{82.5} & \textbf{14} \\
\bottomrule
\end{tabular}
}
\end{table}

\begin{table}[H]
\centering
\caption{Comparison of different methods at a fixed query fraction of $B/N=20\%$ with various watermarking schemes. All results are evaluated using Mistral-7B-Instruct-v0.3 with C4.}
\label{tab:6}
\vspace{4pt}
\resizebox{\linewidth}{!}{%
\begin{tabular}{@{}l*{10}{c}@{}}
\toprule
\multirow{2}{*}{Method} & \multicolumn{5}{c}{KGW} & \multicolumn{5}{c}{Hu Unbiased} \\
 & Prec.\,$\uparrow$ & Rec.\,$\uparrow$ & F1\,$\uparrow$ & IoU\,$\uparrow$ & $B^{\star}(0.8)\,\downarrow$ & Prec.\,$\uparrow$ & Rec.\,$\uparrow$ & F1\,$\uparrow$ & IoU\,$\uparrow$ & $B^{\star}(0.8)\,\downarrow$ \\
\midrule
WaterSeeker & 83.5 & 71.2 & 76.9 & 62.5 & 41 & 83.5 & 63.6 & 72.2 & 56.5 & 47 \\
WinMax & 87.1 & 72.0 & 78.8 & 65.0 & 37 & 85.5 & 70.7 & 77.4 & 63.1 & 42 \\
AOL & 84.6 & 71.2 & 77.3 & 63.0 & 43 & 85.0 & 60.3 & 70.6 & 54.6 & 48 \\
\midrule
\textbf{BELL (ours)} & \textbf{95.8} & \textbf{90.5} & \textbf{93.1} & \textbf{87.0} & \textbf{12} & \textbf{94.2} & \textbf{89.3} & \textbf{91.7} & \textbf{84.6} & \textbf{13} \\
\bottomrule
\addlinespace[7pt]
\toprule
\multirow{2}{*}{Method} & \multicolumn{5}{c}{SynthID-style} & \multicolumn{5}{c}{Gumbel} \\
 & Prec.\,$\uparrow$ & Rec.\,$\uparrow$ & F1\,$\uparrow$ & IoU\,$\uparrow$ & $B^{\star}(0.8)\,\downarrow$ & Prec.\,$\uparrow$ & Rec.\,$\uparrow$ & F1\,$\uparrow$ & IoU\,$\uparrow$ & $B^{\star}(0.8)\,\downarrow$ \\
\midrule
WaterSeeker & 82.8 & 67.2 & 74.2 & 59.0 & 45 & 84.0 & 60.7 & 70.5 & 54.4 & 48 \\
WinMax & 86.3 & 70.3 & 77.5 & 63.3 & 40 & 85.0 & 67.5 & 75.3 & 60.4 & 44 \\
AOL & 84.5 & 64.7 & 73.3 & 57.9 & 47 & 84.9 & 57.9 & 68.9 & 52.6 & 50 \\
\midrule
\textbf{BELL (ours)} & \textbf{95.3} & \textbf{89.5} & \textbf{92.3} & \textbf{85.7} & \textbf{12} & \textbf{94.7} & \textbf{89.7} & \textbf{92.1} & \textbf{85.4} & \textbf{12} \\
\bottomrule
\end{tabular}
}
\end{table}

\begin{table}[H]
\centering
\caption{Comparison of different methods at a fixed query fraction of $B/N=20\%$ with various watermarking schemes. All results are evaluated using Mistral-7B-Instruct-v0.3 with WikiText-103.}
\label{tab:7}
\vspace{4pt}
\resizebox{\linewidth}{!}{%
\begin{tabular}{@{}l*{10}{c}@{}}
\toprule
\multirow{2}{*}{Method} & \multicolumn{5}{c}{KGW} & \multicolumn{5}{c}{Hu Unbiased} \\
 & Prec.\,$\uparrow$ & Rec.\,$\uparrow$ & F1\,$\uparrow$ & IoU\,$\uparrow$ & $B^{\star}(0.8)\,\downarrow$ & Prec.\,$\uparrow$ & Rec.\,$\uparrow$ & F1\,$\uparrow$ & IoU\,$\uparrow$ & $B^{\star}(0.8)\,\downarrow$ \\
\midrule
WaterSeeker & 83.8 & 69.7 & 76.1 & 61.4 & 42 & 82.3 & 61.5 & 70.4 & 54.3 & 48 \\
WinMax & 85.9 & 71.8 & 78.2 & 64.2 & 38 & 86.0 & 66.1 & 74.8 & 59.7 & 43 \\
AOL & 85.6 & 66.7 & 75.0 & 60.0 & 44 & 83.4 & 59.4 & 69.4 & 53.1 & 51 \\
\midrule
\textbf{BELL (ours)} & \textbf{95.2} & \textbf{89.5} & \textbf{92.3} & \textbf{85.6} & \textbf{12} & \textbf{94.6} & \textbf{87.6} & \textbf{91.0} & \textbf{83.4} & \textbf{14} \\
\bottomrule
\addlinespace[7pt]
\toprule
\multirow{2}{*}{Method} & \multicolumn{5}{c}{SynthID-style} & \multicolumn{5}{c}{Gumbel} \\
 & Prec.\,$\uparrow$ & Rec.\,$\uparrow$ & F1\,$\uparrow$ & IoU\,$\uparrow$ & $B^{\star}(0.8)\,\downarrow$ & Prec.\,$\uparrow$ & Rec.\,$\uparrow$ & F1\,$\uparrow$ & IoU\,$\uparrow$ & $B^{\star}(0.8)\,\downarrow$ \\
\midrule
WaterSeeker & 83.5 & 63.2 & 72.0 & 56.3 & 46 & 81.9 & 58.7 & 68.4 & 52.0 & 51 \\
WinMax & 85.5 & 70.8 & 77.5 & 63.3 & 41 & 86.3 & 63.4 & 73.1 & 57.6 & 45 \\
AOL & 84.7 & 60.7 & 70.7 & 54.7 & 48 & 84.1 & 56.9 & 67.9 & 51.4 & 53 \\
\midrule
\textbf{BELL (ours)} & \textbf{94.4} & \textbf{88.7} & \textbf{91.5} & \textbf{84.3} & \textbf{13} & \textbf{94.9} & \textbf{88.2} & \textbf{91.4} & \textbf{84.2} & \textbf{13} \\
\bottomrule
\end{tabular}
}
\end{table}

Across the remaining model--dataset configurations, BELL consistently maintains high localization accuracy while requiring substantially fewer detector evaluations to reach the target recall. In particular, $B^\star(0.8)$ ranges from only 12\% to 15\% for BELL across the four watermark families, whereas the competing methods generally require considerably larger query fractions. The improvements are also consistent in Precision, Recall, F1, and IoU, indicating that the query efficiency of BELL does not come at the expense of localization quality. These results complement Table 1 in the main text and show that the observed performance is consistent across both LLM backbones and source datasets.

\subsection{Explainability Experiment}
To further understand why BELL achieves higher query efficiency, we extend the explainability analysis in Figure 2 to the remaining model--dataset configurations. We consider two complementary metrics introduced in Section 5.3. The Target Query Ratio (TQR) measures the fraction of queries allocated to ground-truth watermarked regions, reflecting whether the query budget is concentrated on relevant blocks. Evidence Gain measures the average positive log-evidence obtained per query, characterizing how much useful watermark evidence is extracted from each detector evaluation. Higher values of these two metrics therefore indicate more targeted query allocation and more effective evidence acquisition, respectively.

\figpaperhere{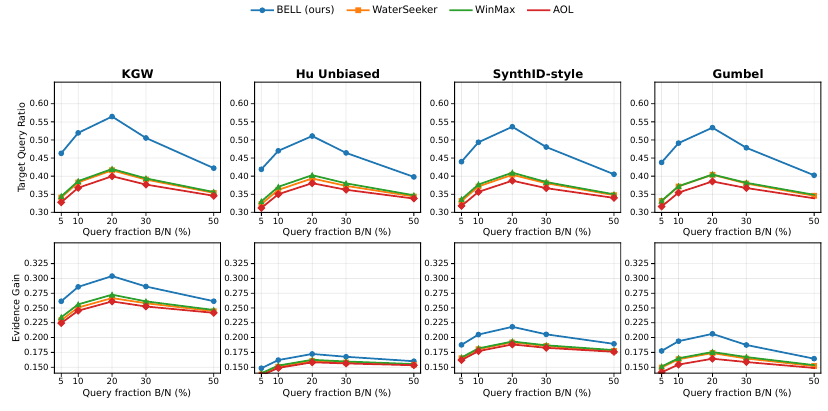}{1}{Evaluation of different methods' query efficiency through the target query ratio (TQR) and evidence gain (EG). All results are evaluated using Qwen2.5-7B-Instruct with WikiText-103.}{fig:6}
\figpaperhere{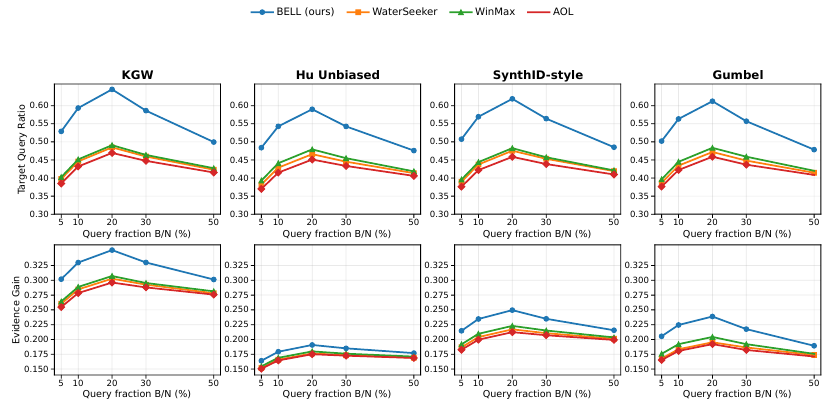}{1}{Evaluation of different methods' query efficiency through the target query ratio (TQR) and evidence gain (EG). All results are evaluated using Mistral-7B-Instruct-v0.3 with C4.}{fig:7}
\figpaperhere{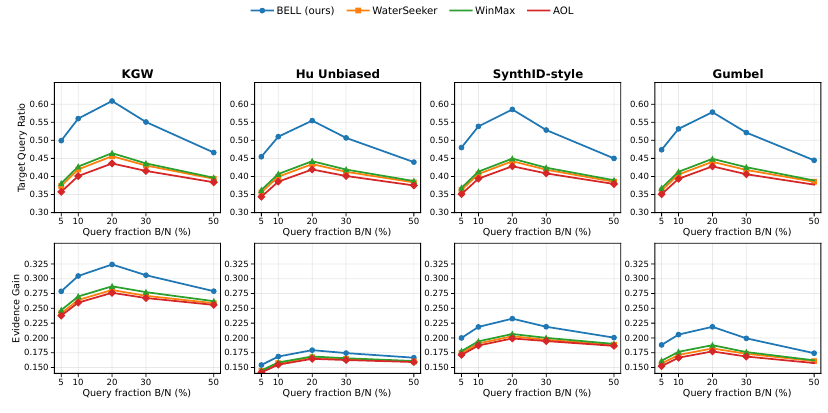}{1}{Evaluation of different methods' query efficiency through the target query ratio (TQR) and evidence gain (EG). All results are evaluated using Mistral-7B-Instruct-v0.3 with WikiText-103.}{fig:8}

Across all three additional configurations, BELL consistently achieves higher Target Query Ratio and Evidence Gain than WaterSeeker, WinMax, and AOL across the four watermark families. Both metrics generally increase as the query fraction grows from 5\% to around 20\%, indicating that BELL progressively concentrates its budget on watermarked regions while accumulating increasingly informative evidence. After this point, the metrics gradually decrease as additional queries are extended to less informative regions. The consistent behavior across different LLM backbones and source datasets supports the same explanation observed in Figure 2: BELL improves query efficiency by both querying the right regions and extracting more useful evidence from each detector evaluation.

\clearpage
\subsection{The Ablation Results}
We report the complete ablation results for the remaining model--dataset configurations at a fixed query fraction of $B/N=20\%$. Following the same setting as Table 2 in the main text, we remove one component at a time from BELL, including coverage, optimistic exploration, neighborhood expansion, and near-confirmation. We report Recall, IoU, and Evidence Gain (EG) under the four watermark families: KGW, Hu Unbiased, SynthID-style, and Gumbel. All results are reported as percentages, and higher values indicate better performance.
\begin{table}[H]
\centering
\caption{Comparison of different methods at a fixed query fraction of $B/N=20\%$ with various watermarking schemes. All results are evaluated using Qwen2.5-7B-Instruct with WikiText-103.}
\label{tab:8}
\vspace{4pt}
\resizebox{\linewidth}{!}{%
\begin{tabular}{@{}l*{12}{c}@{}}
\toprule
\multirow{2}{*}{Variant} & \multicolumn{3}{c}{KGW} & \multicolumn{3}{c}{Hu Unbiased} & \multicolumn{3}{c}{SynthID-style} & \multicolumn{3}{c}{Gumbel} \\
\cmidrule(lr){2-4}\cmidrule(lr){5-7}\cmidrule(lr){8-10}\cmidrule(lr){11-13}
 & Recall & IoU & EG & Recall & IoU & EG & Recall & IoU & EG & Recall & IoU & EG \\
\midrule
w/o Coverage & 82.0 & 76.4 & 28.5 & 75.7 & 74.3 & 14.3 & 79.7 & 73.4 & 20.0 & 76.8 & 75.7 & 17.5 \\
w/o Optimism & 82.6 & 81.0 & 26.1 & 82.2 & 75.9 & 16.9 & 80.9 & 78.8 & 18.5 & 83.6 & 77.1 & 19.9 \\
w/o Neighbor & 83.5 & 77.9 & 29.6 & 77.9 & 76.4 & 14.8 & 82.1 & 76.2 & 21.8 & 79.4 & 77.6 & 17.9 \\
w/o Confirmation & 83.3 & 81.7 & 26.9 & 80.8 & 79.0 & 13.4 & 84.7 & 77.7 & 18.8 & 81.5 & 79.9 & 15.9 \\
\midrule
\textbf{Full BELL} & \textbf{88.6} & \textbf{84.7} & \textbf{30.4} & \textbf{87.1} & \textbf{82.7} & \textbf{17.2} & \textbf{88.1} & \textbf{83.6} & \textbf{21.8} & \textbf{86.6} & \textbf{82.5} & \textbf{20.6} \\
\bottomrule
\end{tabular}
}
\end{table}

\begin{table}[H]
\centering
\caption{Comparison of different methods at a fixed query fraction of $B/N=20\%$ with various watermarking schemes. All results are evaluated using Mistral-7B-Instruct-v0.3 with C4.}
\label{tab:9}
\vspace{4pt}
\resizebox{\linewidth}{!}{%
\begin{tabular}{@{}l*{12}{c}@{}}
\toprule
\multirow{2}{*}{Variant} & \multicolumn{3}{c}{KGW} & \multicolumn{3}{c}{Hu Unbiased} & \multicolumn{3}{c}{SynthID-style} & \multicolumn{3}{c}{Gumbel} \\
\cmidrule(lr){2-4}\cmidrule(lr){5-7}\cmidrule(lr){8-10}\cmidrule(lr){11-13}
 & Recall & IoU & EG & Recall & IoU & EG & Recall & IoU & EG & Recall & IoU & EG \\
\midrule
w/o Coverage & 79.6 & 78.6 & 29.3 & 79.6 & 73.9 & 18.3 & 79.8 & 78.0 & 20.6 & 82.7 & 76.5 & 22.2 \\
w/o Optimism & 86.7 & 81.0 & 33.5 & 82.5 & 80.6 & 16.3 & 85.4 & 79.3 & 24.0 & 83.1 & 81.7 & 20.5 \\
w/o Neighbor & 82.5 & 81.4 & 31.3 & 83.5 & 77.5 & 18.7 & 82.1 & 80.5 & 21.5 & 84.1 & 78.2 & 23.6 \\
w/o Confirmation & 88.0 & 81.9 & 29.6 & 83.5 & 81.9 & 14.8 & 87.5 & 80.9 & 21.3 & 83.5 & 82.2 & 18.9 \\
\midrule
\textbf{Full BELL} & \textbf{90.5} & \textbf{87.0} & \textbf{35.1} & \textbf{89.3} & \textbf{84.6} & \textbf{19.1} & \textbf{89.5} & \textbf{85.7} & \textbf{25.0} & \textbf{89.7} & \textbf{85.4} & \textbf{23.9} \\
\bottomrule
\end{tabular}
}
\end{table}

\begin{table}[H]
\centering
\caption{Comparison of different methods at a fixed query fraction of $B/N=20\%$ with various watermarking schemes. All results are evaluated using Mistral-7B-Instruct-v0.3 with WikiText-103.}
\label{tab:10}
\vspace{4pt}
\resizebox{\linewidth}{!}{%
\begin{tabular}{@{}l*{12}{c}@{}}
\toprule
\multirow{2}{*}{Variant} & \multicolumn{3}{c}{KGW} & \multicolumn{3}{c}{Hu Unbiased} & \multicolumn{3}{c}{SynthID-style} & \multicolumn{3}{c}{Gumbel} \\
\cmidrule(lr){2-4}\cmidrule(lr){5-7}\cmidrule(lr){8-10}\cmidrule(lr){11-13}
 & Recall & IoU & EG & Recall & IoU & EG & Recall & IoU & EG & Recall & IoU & EG \\
\midrule
w/o Coverage & 82.1 & 76.6 & 30.1 & 76.2 & 74.6 & 14.7 & 80.6 & 74.8 & 22.1 & 78.0 & 76.2 & 17.8 \\
w/o Optimism & 83.2 & 81.7 & 27.7 & 83.3 & 77.1 & 17.6 & 82.2 & 80.2 & 19.7 & 84.2 & 78.2 & 21.0 \\
w/o Neighbor & 84.2 & 78.6 & 31.4 & 80.1 & 77.9 & 15.3 & 83.6 & 77.5 & 22.7 & 80.5 & 78.7 & 19.1 \\
w/o Confirmation & 84.5 & 83.3 & 25.7 & 84.6 & 78.0 & 15.5 & 83.2 & 81.6 & 18.4 & 85.3 & 78.7 & 18.9 \\
\midrule
\textbf{Full BELL} & \textbf{89.5} & \textbf{85.6} & \textbf{32.4} & \textbf{87.6} & \textbf{83.4} & \textbf{17.9} & \textbf{88.7} & \textbf{84.3} & \textbf{23.3} & \textbf{88.2} & \textbf{84.2} & \textbf{21.9} \\
\bottomrule
\end{tabular}
}
\end{table}

Across the three additional model--dataset configurations, the ablation results exhibit a consistent pattern with the results reported in the main text. Removing coverage leads to the largest degradation in Recall and IoU across the four watermark families, highlighting the importance of maintaining broad exploration before concentrating the query budget. Neighborhood expansion also consistently improves localization quality by leveraging the spatial continuity of watermarked regions. In contrast, removing near-confirmation has a relatively smaller effect on Recall and IoU but results in a more pronounced reduction in Evidence Gain, indicating that near-confirmation primarily improves the efficiency of evidence acquisition.

\end{document}